\documentclass[nofootinbib,pra,onecolumn,showpacs,notitlepage,amsmath,amssymb,superscriptaddress]{revtex4-1} %

\usepackage[colorlinks=true,citecolor=blue,linkcolor=blue,urlcolor=blue]{hyperref}
\usepackage{cleveref}[2012/02/15]
\crefformat{footnote}{#2\footnotemark[#1]#3}

\usepackage{amsmath,amssymb,bm}

\usepackage{tikz}
\usetikzlibrary{external}
\usetikzlibrary{decorations.pathreplacing}
\usetikzlibrary{decorations.text}
\usetikzlibrary{decorations.pathmorphing}
\usetikzlibrary{shapes}

\definecolor{newblue}{RGB}{112,178,255}
\definecolor{neworange}{RGB}{255,204,112}
\definecolor{blue2}{RGB}{120,0,255}
\definecolor{red2}{RGB}{255,0,120}
\definecolor{green2}{RGB}{0,130,130}

\tikzset{snake it/.style={decorate, decoration={snake,segment length=1mm, amplitude=0.5mm}}}

\definecolor{darkred}{RGB}{245,186,183}
\definecolor{lightred}{RGB}{249,217,215}

\usetikzlibrary{arrows,calc,external,shapes.geometric}

\tikzset{
>=stealth',
help lines/.style={dashed, thick},
important line/.style={thick},
connection/.style={thick, dotted},
}

\tikzstyle{A}=[circle,draw=red!50,fill=red!20,thick]
\tikzstyle{R}=[circle,draw=blue!50,fill=blue!20,thick]
\tikzstyle{U}=[circle,draw=green!50,fill=green!20,thick]
\tikzstyle{V}=[circle,draw=orange!50,fill=orange!20,thick]

\tikzstyle{bag} = [align=center]

\usepackage{tikz-3dplot}

\def\bra#1{\mathinner{\langle{#1}|}}
\def\ket#1{\mathinner{|{#1}\rangle}}

\newcommand{\eeq}{\end{equation}}
\newcommand{\ea}{\end{array}}

\def\eea{\end{eqnarray}}

\def\<{\langle}
\def\>{\rangle}

\usepackage{amsmath}
\usepackage{comment}
\usepackage{amssymb}
\usepackage{amsthm}

\theoremstyle{definition}

\makeatletter
\renewcommand\onecolumngrid{
\do@columngrid{one}{\@ne}%
\def\set@footnotewidth{\onecolumngrid}
\def\footnoterule{\kern-6pt\hrule width 1.5in\kern6pt}%
}
\makeatother

\begin{document}

\title{Multicriticality between Purely Gapless SPT Phases with Unitary Symmetry}

\author{Saranesh Prembabu}
\affiliation{Department of Physics, Harvard University, Cambridge, MA 02138, USA}

\author{Ruben Verresen}
\affiliation{Pritzker School of Molecular Engineering, University of Chicago, Chicago, IL 60637, USA}

\begin{abstract}
Symmetry-protected topological (SPT) phases are commonly required to have an energy gap, but recent work has extended the concept to gapless settings. This raises a natural question: what happens at transitions between inequivalent gapless SPTs? We address this for the simplest known case among gapless SPTs protected by a unitary symmetry group acting faithfully on the low-energy theory. To this end, we consider a qutrit version of the nearest-neighbor XX chain. Trimerizing the chain explicitly breaks an anomalous symmetry and produces three distinct gapped SPT phases protected by a unitary $\mathbb{Z}_3 \times \mathbb{Z}_3$ symmetry. Their phase boundaries are given by three inequivalent gapless SPTs without any gapped symmetry sectors, each described by a symmetry-enriched version of an orbifolded Potts$^2$ conformal field theory with central charge $c=\frac{8}{5}$. We provide an analytic derivation of this critical theory in a particular regime and confirm its stability using tensor network simulations. Remarkably, the three gapless SPTs meet at a $c = 2$ multicritical point, where the protecting $\mathbb{Z}_3 \times \mathbb{Z}_3$ symmetry exhibits a mixed anomaly with the $\mathbb Z_3$ entangler symmetry that permutes the SPT classes. We further explore how discrete gauging gives dipole-symmetric models, offering insights into dipole symmetry-breaking and SPTs, as well as symmetry-enriched multiversality. Altogether, this work uncovers a rich phase diagram of a minimal qutrit chain, whose purely nearest-neighbor interactions make it a promising candidate for experimental realization, including the prospect of critical phases with stable edge modes.
\end{abstract}

\date{\today}

\maketitle

\tableofcontents

\section{Introduction}

\label{sec:intro}

Over the past two decades, \emph{symmetry-protected topological} (SPT) phases have become a central theme in condensed-matter physics and quantum information \cite{Gu_2009, Pollmann_2010, Pollmann_2012,  Turner_2011, Fidkowski_2011, Schuch_2011, Chen_2011, Chen_2011_2, Chen_2013, Chen2012SPT, Else_2014, Wang_2014, Senthil_2015, Kapustin_2015}. In their original gapped incarnation, a nontrivial SPT phase is a short-range–entangled many-body state that is adiabatically connected to a trivial product state \emph{if and only if} the protecting symmetry is broken. Such nontrivial SPTs exhibit non-trivial features such as robust gapless edge modes---at least for internal symmetries which will be the focus of the present work.
For an on-site (i.e., tensor product) unitary symmetry group $G$, the equivalence classes of \emph{gapped} bosonic SPTs are well understood; for example in low spatial dimensions $d$ they are classified by the cohomology group $H^{d+1}(G,U(1))$~\cite{Chen_2013}. In this work our focus is on the one-dimensional case ($d=1$). 
The non-trivial phenomenology of an SPT phase is revealed most clearly when \emph{two inequivalent SPTs} appear in a phase diagram. Because they cannot be connected without breaking $G$ or closing the energy gap above the ground state, one is forced to consider the \emph{phase transition} between them \cite{Jiang_2010, Nonne_2013, Morimoto_2014, Alet_2011, Smacchia_2011,  Liu_2011, Ueda_2014, Chen_2015, Lecheminant_2015, Chen_2013_transition, Duivenvoorden13a, DeGottardi,Lahtinen_2015,  Nataf_2016, Prakash_2016, Rao_2016, Tsui_2015,Tsui_2017, Furaya_2017, Verresen17,Verresen_2018, Roy_2018, Prembabu_2022,cordova2022symmetryenrichedctheorems, Antinucci25}. 

\subsection{Gapless SPT phases}

The SPT concept has more recently been extended to \emph{gapless} systems, often described at low energies by conformal field theories (CFTs) \cite{Scaffidi_2017,Verresen_2021}. We say two Hamiltonians $H_0$ and $H_1$ are distinct \emph{symmetry-enriched criticalities} (SEC) \cite{Verresen_2021} if they are described by the same CFT but they \textit{cannot} be continuously deformed into one another along a path of local $G$-symmetric Hamiltonians $H_\lambda$ which is described by the same CFT\footnote{More succinctly, $H_0$ and $H_1$ are described by the 
\emph{same} CFT if we ignore the (UV) symmetry group $G$, but they are in \emph{distinct} $G$-CFTs where we enforce the symmetry---hence the term `symmetry-enriched CFT' \cite{Verresen_2021}. This is analogous to the well-explored concept of symmetry-enriched topological order \cite{Mesaros_2013, Barkeshli_2019}.} for all $0 \leq \lambda \leq 1$.  A \emph{gapless SPT} (gSPT) is then a particularly sharp kind of SEC, in which even the local operator content's symmetry charges coincide, so that no bulk local probe distinguishes the two phases. They may be distinguished by the charges of nonlocal observables, boundaries or twisted sectors. Distinct gSPTs built from the same CFT data therefore represent distinct quantum phases protected by $G$. While the symmetry $G$ does not necessarily rule out relevant perturbations gapping out of the CFT (i.e., we do not require that $G$ stabilizes the gapesslessness), it does rule out continuous Hamiltonian paths remaining within the universality class from one gSPT to another.

Concrete lattice examples already span a variety of symmetries and CFTs~\cite{Kestner11,Grover12,Keselman_2015,Scaffidi_2017,Parker_2018,Verresen_2018,Verresen_2021,Thorngren_2021,Borla_2021,Thorngren_2021, Duque_2021,Hidaka_2022,Prembabu_2022,Ma_2022, Wen_2023, wen2023classification11dgaplesssymmetry,Li_2024,Lei24,Yu24,Ando24,bhardwaj2024hassediagramsgaplessspt,qi2025symmetrytacoequivalencesgapped,Linhao25,Calderon25,wen2025topologicalholography21dgapped,Antinucci25,Rey_2025}. A particular subclass---which we focus on in the present work---are ``purely gapless'' SPTs, by which we mean there are no gapped symmetry sectors involved. To the best of our knowledge, the simplest such example with a unitary symmetry group is the $c=8/5$ orbifold Potts$^{2}$ gSPT protected by $\mathbb Z_{3}\!\times\!\mathbb Z_{3}$ ~\cite{Prembabu_2022}.  Incorporating anti-unitary symmetries allows further examples, such as the \emph{Ising$^{\ast}$} chain protected by $\mathbb Z_{2}\!\times\!\mathbb Z_{2}^{\mathcal T}$, which hosts a time-reversal–odd disorder operator and algebraically localized edge modes \cite{Verresen_2021}. Beyond such ``pure'' gSPTs, there are also cases where a subgroup of the microscopic symmetry is gapped out and only a quotient acts faithfully at low energies \cite{Scaffidi_2017}. These include all known examples of intrinsically gapless SPTs (igSPTs), where the emergent low-energy quotient symmetry itself is anomalous \cite{Thorngren_2021}.

\subsection{Multicriticality of gapless SPT phases}
The interplay of inequivalent gSPTs with the same operator content shows promise for novel physics. It naturally raises the question: \emph{how can one tune between them while preserving the protecting symmetry}?  Such tuning can, in principle, proceed through a \emph{multicritical point} described by a richer CFT (with higher central charge in 1+1d), directly analogous to a second-order transition between gapped SPTs. Such a multicritical CFT would admit different relevant perturbations flowing into the distinct gapless SPTs. 

Here we construct a nearest-neighbor \emph{qutrit} spin chain protected by the fully unitary symmetry $G=\mathbb Z_{3}\times\mathbb Z_{3}$ and show that it hosts three inequivalent gSPT phases, each with central charge $c=\tfrac85$, that collide at a single multicritical point with $c=2$.  
This constitutes the first example of a multicritical transition between inequivalent gapless SPTs in which the protecting symmetry is \emph{unitary} and acts \emph{faithfully} on the low‐energy theory. Moreover, these gapless SPTs all have central charges larger than one. All couplings are nearest-neighbor, making them experimentally feasible, such as potentially in neutral-atom quantum simulators~\cite{de_L_s_leuc_2019, Mishmash_2011, Capponi_2020, Fromholz_2019, Sule_2015}.  

A handful of examples of gSPT transitions are known in other settings with symmetries acting either anti-unitarily or non-faithfully on the low energy theory (i.e., certain symmetries act on gapped sectors). 
For instance, the Ising $\leftrightarrow$ Ising$^{\ast}$ multicriticality can realize either a $c=1$ Gaussian theory or a $z=2$ Lifshitz point~\cite{Verresen_2018,Yu_2024}, while an interacting-fermion igSPT exhibits a $c=\tfrac32$ transition~\cite{Zhang_2024}.  
In spin-1 chains, a Lifshitz point connects the trivial Ising, nontrivial Ising, and an igSPT phase \cite{Yang_2023}, although the protecing symmetries act non-faithfully on the CFTs. 
Distinct $c=1$ gSPTs with $\mathbb Z_{2}\times\mathbb Z_{2}\times\mathbb Z_{2}^{\mathcal T}$ symmetry were  found to connect through a broad incommensurate ferromagnetic region rather than a single multicritical point \cite{Rey_2025}.  
Related transitions also appear for locally symmetry-enriched criticalities, such as between inequivalent Luttinger liquids \cite{Mondal_2023}.

\subsection{Entangler and mixed anomaly}
\label{subsec:entangler}

As a useful guide to constructing new models, we focus on gapless SPTs related by an \emph{entangler}.  An entangler \(U\) is a locality-preserving unitary that toggles between distinct gapped \(G\)-SPT phases~\cite{Chen_2013, Chen_2014,Else_2014, Bultinck_2019, pivot, zhang2022topologicalinvariantssptentanglers}. We typically require that $U$ is globally $G$-symmetric, such that it does not change the charge of (untwisted) local operators, but multiplies the charge of nonlocal \(g\)-string operators by a fixed phase \(e^{i\alpha(g)}\) (with \(\alpha(1)=0\)). (Equivalently, \(U\) acts on the low energy theory by stacking a \(G\)-SPT). 
Gapless SPTs can be said to be related by an entangler if they have isomorphic operator content, but the charges of the $g$-twisted sector are shifted by \(e^{i\alpha(g)}\) in one relative to the other. The entangler toggles between distinct gapless SPTs by implementing this charge shift, and is thus \textit{not} a symmetry of the CFT. Such gapless SPTs naturally arise at second order transitions of gapped SPTs (see e.g. Refs.~\onlinecite*{Verresen_2021,Prembabu_2022}). 
We note that since by definition such gapless SPTs can be toggled by stacking with gapped SPT phases, they \emph{cannot} be intrinsically gapless SPT phases~\cite{Thorngren_2021}. Nevertheless, they can be \emph{purely gapless} if all the symmetry sectors involved in the gapped SPT also act faithfully on the low-energy gapless theory. (We note that currently there are no known examples of SPTs which are purely gapless \emph{and} intrinsically gapless.)

Now consider a point in parameter space where the gSPTs related by entanglers meet. At such a multicritical point, it is natural for \(U\) to act as a symmetry of the low-energy theory. As a symmetry it would shift the charges of $G$-twisted sectors. Thus we would expect the multicritical theory to have a mixed anomaly between $U$ and $G$, given by a type-III cocycle $\omega_3 \in H^{d+2}(U\times G, U(1))$. This is the gapless analogue of familiar arguments about anomalies at gapped SPT transitions \cite{Bultinck_2019,Tsui_2015,Lanzetta_2023}. The anomalous symmetry may be emergent in the infrared. In the example we study in this paper, the entangler symmetry is exact on the lattice (i.e., single-site translation) so the anomaly is explicit.

\section{Hamiltonian Model}

We can consider the minimal known case of a unitary symmetry hosting multiple purely gapless SPT phases related by an SPT entangler, namely the $G =\mathbb{Z}_3 \times \mathbb{Z}_3$ symmetry group. The entangler of the SPTs for this group acts by a $\mathbb{Z}_3$ action.

A naive approach to look for a multicritical point is to linearly interpolate between the three known solvable qutrit ``cluster Hamiltonians'' \cite{Briegel_2001} for the $\mathbb{Z}_3\times \mathbb{Z}_3$ SPT phases, as defined in Refs.~\onlinecite{geraedts2014exactmodelssymmetryprotectedtopological,Santos15}. Unfortunately this does not work. Although interpolating between two of cluster Hamiltonians does result in a gapless SPT \cite{Tsui_2017,Prembabu_2022} (namely, an orbifold Potts$^2$ $c=8/5$ theory), the third cluster Hamiltonian acts as a relevant perturbation that immediately gaps out the theory. With positive linear combinations of all three cluster Hamiltonians, there is no second order SPT transition, and there is an intermediate symmetry-breaking phase~\cite{Bibo_Verresen_Pollmann_toappear}. Thus a more principled strategic approach is needed, using different microscopic realizations.

Our strategy will be to work backwards.
As we had argued earlier, a candidate CFT for multicriticality would likely feature a mixed anomaly between the protecting group $G = \mathbb{Z}_3 \times \mathbb{Z}_3$ and the SPT entangler group $\mathbb{Z}_3$. We will thus consider known CFTs with this $\mathbb{Z}_3 \times \mathbb{Z}_3 \times\mathbb{Z}_3$ anomalous symmetry for some group realization. Then we can explore explicit breaking of the entangler group symmetry with the hope of seeing an RG flow into a gapless SPT. While such a flow can be very challenging to predict, we can be aided if we can prove that we obtain the correct CFT for strong perturbation, supplemented by numerical checks at intermediate perturbation strength.

The above considerations lead us to a promising candidate for a multicritical transition, in particular the spatially-modulated two-body qutrit XXYY model (and its special case, the qutrit XX model), which which we will introduce below. This is a gapless qutrit model with nearest neighbour interactions and anomalous $\mathbb{Z}_3 \times \mathbb{Z}_3 \times \mathbb{Z}_3
$ symmetry action, as well as deformations breaking part of the anomalous symmetry action, enabling flows to gapped and gapless SPT phases. To construct this model, we will be guided by a well-known analogous situation for a qubit model.

\subsection{Warm-up: bond-alternating qubit XX chain}

To set the stage for our main discussion on $\mathbb{Z}_3\times \mathbb{Z}_3$ SPT phases, it is instructive to review the more familiar example of the transition between gapped $\mathbb{Z}_2 \times \mathbb{Z}_2$ phases in the bond-alternating XX model. This will illustrate what we need to know about the concept of a Lieb-Schultz-Mattis (LSM) anomaly \cite{Lieb_Schultz_Mattis_1961, Oshikawa_2000, Hastings_2004, Hastings_2005} in the context of gapped SPT phases for which translation is the entangler.

On a one-dimensional chain of qubits, the XX model is given by
\begin{equation}
H^{\textrm{qubit}}_{XX}  = \sum_j  a_{j\bmod 2} \left( \sigma^x_j \sigma^x_{j+1}+\sigma^y_j \sigma^y_{j+1}
+ \textrm{h.c.}\right), \label{eq:qubitXX}
\end{equation}
where $a_0, a_1 \ge 0$ are real coefficients. The model respects a $\mathbb{Z}_2 \times \mathbb{Z}_2$ symmetry generated by $\prod_j \sigma^x_j$ and $\prod_j \sigma^y_j$, realized projectively on-site. Upon Jordan-Wigner transformation, this model is also equivalent to the Su-Schrieffer-Heeger (SSH) model of free fermions~\cite{SSH}.
As an aside, unlike the canonical example of the cluster chain SPT, this Hamiltonian has strictly nearest-neighbor two-body interations, which has allowed it to be experimentally realized in Rydberg atom tweezer arrays~\cite{de_L_s_leuc_2019}. However, there is a simple local unitary mapping to the (perturbed) cluster chain \cite{Verresen17}; this will not be the case for the qutrit case which we will discuss soon.

\begin{figure}
\centering
\includegraphics[scale=0.4]
{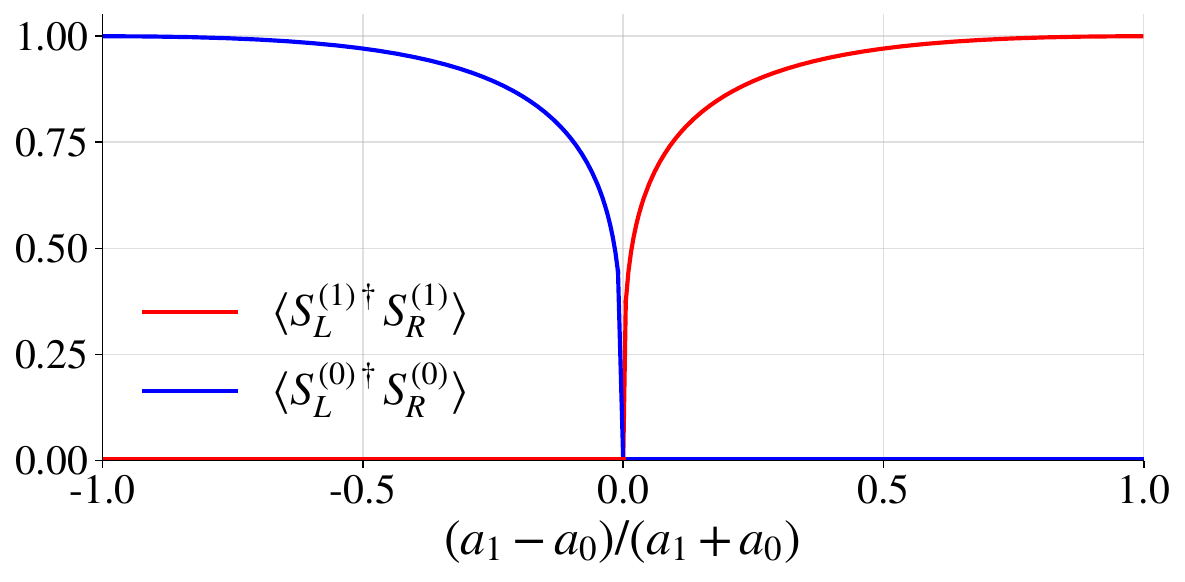}
\caption{\textbf{Long-range string order parameters in the `warm-up' qubit model.} We consider the bond-alternating XX chain in Eq.~\eqref{eq:qubitXX}, which gives a free-fermion solvable SPT transition. We show the asymptotic values of the string order parameters \eqref{eq:qubitSOP} which distinguish the two gapped SPT phases protected by $\mathbb Z_2^X \times \mathbb Z_2^Z$ symmetry. At the translation-invariant point $a_0=a_1$, the string order parameters are treated on an equal footing and thus forbid a gapped nondegenerate phase; this is a manifestation of the LSM anomaly. In this case, the theory with the LSM anomaly is a $c=1$ conformal field theory (at the free-fermion point \cite{Ginsparg}), which implies that the string order parameters vanish with a universal dependence $\sim |a_1-a_0|^{1/4}$. }
\label{fig:qubitXX}
\end{figure}

For $a_0 > a_1$ and $a_0<a_1$ this model realizes two distinct SPT phases \cite{Verresen_2021}, each of which can be diagnosed by string order parameters 
\begin{equation}
S^{(0)}_j = \prod_{k \leq 2j} \sigma^x_k,
\qquad
S^{(1)}_j = \prod_{k \leq 2j+1} \sigma^x_{k}.
\label{eq:qubitSOP}
\end{equation}
Observe that these string order parameters have different charges under the global $\mathbb{Z}_2$ symmetry subgroup generated by $\prod_{k \in \mathbb{Z}} \sigma^y_k$. Although the precise value of the charge depends on boundary conditions, it is clear that there is a relative difference of $-1$ between the charges of the two string order parameters (indeed, $S^{(1)}_j = S^{(0)}_j \times \sigma^x_{2j+1}$). For each SPT phase, exactly \textit{one} of these two string order parameters has long range order, $\lim_{|i-j| \to \infty} \langle S_i^{(k)\dagger} S_j^{(k)} \rangle \neq 0$, and the other does not, as in Figure \ref{fig:qubitXX}. In the extreme limits $a_0 = 0$ or $a_1 = 0$, the chain is perfectly dimerized into singlets $\bigotimes \left( \frac{ \ket{01}+ \ket{10}}{\sqrt{2}} \right)$. The nontrivial SPT phase is adiabatically linked to the AKLT Haldane phase \cite{AKLT,Verresen17}.


At the translationally symmetric point $a_0 = a_1$, there is an anomaly between translation and the onsite $\mathbb{Z}_2^X\times \mathbb{Z}_2^Y$ symmetry, known as the LSM anomaly. One way to characterize the anomaly is that it forbids any short-range entangled symmetric ground state (in particular forbidding a unique symmetric ground state in one spatial dimension). This fact is easy to see using the distinct charges of the string order parameters. A nondegenerate gapped phase as a rule can only have one unique charge for a string order parameter\footnote{Otherwise one could multiply the differently-charged string operators and obtain long-range order for a local charged operator, implying a degeneracy.}. The fact that all the differently charged operators are related by translational thus automatically rules out the possibility of a nondegenerate gapped phase.

In this example, which is free-fermion solvable, we see that what results is a gapless theory whose low energy description is a compact boson CFT with central charge $c=1$. Furthermore there is anomaly matching between the microscopic theory and the anomaly of the CFT \cite{Metlitski_2018}. While translation by one site has a mixed anomaly, translation by two sites does not. Thus it is natural to think of the system as having a two-site unit cell. 
The $\mathbb{Z}_2^{\textrm{Trans.}}$ quotient acts as an internal symmetry of the CFT.
The LSM anomaly manifests in the CFT through the $F$ symbols of the topological defect lines corresponding to the translation $\mathbb{Z}_2^{\textrm{Trans.}}$ and global $\mathbb{Z}_2^X\times \mathbb{Z}_2^Y$ symmetries; in the presence of a spatial defect line the temporal defect lines anticommute ~\cite{Chang_2019, thorngren2021fusion}.

The distinction between the SPT phases is manifest on a chain with open boundary conditions. If we consistently terminate the chain onto the interval \(1 \leq j \leq 2N\), then we observe that \(a_0 < a_1\) realizes a ``nontrivial'' phase with edge modes whereas \(a_0 > a_1\) realizes a ``trivial'' phase without edge modes (where we consider the regime $a_0,a_1 \geq 0$). The edge modes transform under a projective symmetry representation on each edge. This can be seen manifestly in the exactly dimerized case where the operators $X$ and $Y$ anticommute on the decoupled mode at the endpoint; the projective $-1$ phase at the endpoint is a universal feature of the nontrivial phase. The notion of which phase is trivial and nontrivial respectively depends on the endpoint termination, but no matter what the endpoint termination is, a robust difference is seen between the two phases’ edge behavior. It is also possible to generalize this to the bond-alternating XXZ model, which exhibits similar physics but tunes through a moduli space of CFTs for different $\Delta$ \cite{Affleck_notes}:
\begin{equation}
H^{\textrm{qubit}}_{XXZ}  = \sum_j  a_{j\bmod 2} \left( \sigma^x_j \sigma^x_{j+1}
+\sigma^y_j \sigma^y_{j+1}
+  \Delta \sigma^z_j \sigma^z_{j+1} \right).
\end{equation}

In summary, the bond-alternating XX model allows us to explore a transition of \textit{gapped} SPT phases. In our work we take this concept further to a \emph{qutrit} chain. There we can furthermore to explore a transition of multiple \emph{gapless} SPT phases, which in turn can be interpreted as phase transitions between gapped SPT phases.

\subsection{Trimerized qutrit XXYY and XX chain}

Inspired by the qubit case, we can construct a nearest-neighbor qutrit model with $\mathbb{Z}_3\times \mathbb{Z}_3$ symmetry and an analogous LSM anomaly.  
The goal is to look for a family of gapless LSM anomalous theories, such that for \textit{some} theory in this family, a translation-breaking perturbation leads to an RG flow into a \textit{gapless} theory.
If that gapless theory does not retain the translation subgroup as an emergent symmetry, then that gapless theory is a gapless SPT, and there exist other flows into different inequivalent gapless SPTs. Thus this would be an example of the sought-after multicritical point.

\medskip  
We consider a one-dimensional lattice with a qutrit degree of freedom on each site, 
and a global symmetry
\(
G \equiv \mathbb{Z}_3^X \times \mathbb{Z}_3^Z ,
\)
generated by
\(
\prod_j X_j\) and \(\prod_j Z_j 
\), products of the standard clock and shift matrices on each lattice site:
\[
X = 
\begin{pmatrix}
0 & 0 & 1 \\
1 & 0 & 0 \\
0 & 1 & 0
\end{pmatrix},
\qquad
Z =
\begin{pmatrix}
1 & 0 & 0 \\
0 & \omega & 0 \\
0 & 0 & \omega^2
\end{pmatrix},
\qquad \omega = e^{2\pi i/3}.
\]

The most general translation-invariant $G$-symmetric nearest-neighbor Hamiltonian looks takes the form
\begin{equation}
H = \sum_{j} \big( J_x X_j^\dagger X_{j+1} + J_y Y_j^\dagger Y_{j+1} + J_z Z_j^\dagger Z_{j+1} + J_w W_j^\dagger W_{j+1} + \text{h.c.} \big), \label{eq:XYZW}
\end{equation}
where  $Y = X^\dagger Z X^\dagger  \propto ZX$ and $W = XZX \propto Z X^\dagger$. One can call this the $XYZW$ qutrit chain, in analogy to the qubit case (since the most general translation-invariant $\mathbb Z_2 \times \mathbb Z_2$-symmetric qubit chain is called the $XYZ$ chain). Manifestly any such system features an LSM anomaly for the same reason as in the qubit case: On a single site, $X$ and $Z$ commute with a $\omega$ phase forming a projective representation of $\mathbb{Z}_3\times \mathbb{Z}_3$.
The $\mathbb{Z}_3^X \times \mathbb{Z}_3^Z$ symmetry acts linearly on a three-site unit cell (i.e., $X_1X_2X_3$ commutes with $Z_1Z_2Z_3$), showing that the mixed anomaly is actually with a quotient group of translations that we can call $\mathbb{Z}_3^{\textrm{Trans.}}$. The three-site unit cell here is analogous to the two-site unit cell for the qubit chain. We note that large parts of the phase diagram of this translation-invariant anomalous spin chain have been explored in Ref.~\onlinecite{Alavirad_2021}.

Our objective is to identify gapless LSM-anomalous theories for which spatial modulation induces renormalization group flows into either other gapless phases or gapped SPT phases. In particular, any continuous transition between distinct gapped SPT phases necessarily passes through a gapless SPT phase, providing a natural target for our search. Thus, gapless phases and gapped SPT phases are the desired outcomes of our analysis, whereas flows into conventional symmetry-breaking phases are not. To sharpen this focus, we impose an additional on-site Hadamard symmetry ($X \to Z \to X^\dagger$). While this symmetry is not itself a protecting symmetry in the SPT sense, it has the useful consequence of excluding spontaneous symmetry-breaking phases that would preserve only a residual $\mathbb{Z}_3$ subgroup. In this way, the Hadamard symmetry serves as a selection principle, narrowing the landscape of possible models and guiding us toward the desired gapless theories. In particular, we go from Eq.~\eqref{eq:XYZW} with its \emph{eight} free parameters (note that $J_x,J_y,J_z,J_w \in \mathbb C$, in contrast to the three-parameter qubit case $J_x,J_y,J_z \in \mathbb R$) to the \emph{two} remaining parameters consistent with Hadamard symmetry, i.e., $J_x = J_z \in \mathbb R$ and $J_y = J_w \in \mathbb R$).

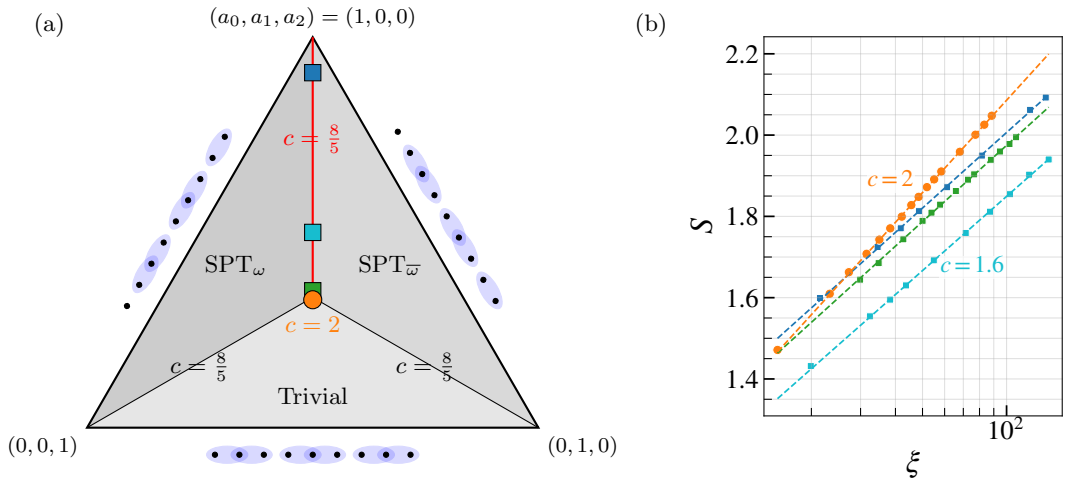
\begin{figure}
\centering
\begin{tikzpicture}
\node at (0,0){
   \begin{tikzpicture}[scale=6]
\definecolor{bluePython}{rgb}{0.12, 0.47, 0.71}  
\definecolor{tealPython}{rgb}{0.09, 0.74, 0.81}  
\definecolor{greenPython}{rgb}{0.17, 0.63, 0.17} 
\definecolor{orangePython}{rgb}{1.0, 0.5, 0.05}  
\definecolor{gray1}{gray}{0.9}
\definecolor{gray2}{gray}{0.85}
\definecolor{gray3}{gray}{0.8}

\coordinate (A) at (0,0);
\coordinate (B) at (1,0);
\coordinate (C) at (0.5,0.866);
\coordinate (O) at (0.5,0.289);

\node[above] at (C) {\footnotesize $(a_0,a_1,a_2) = (1,0,0)$};
\node[below right] at (B) {\footnotesize$(0,1,0)$};
\node[below left] at (A) {\footnotesize$(0,0,1)$};

\fill[gray1] (A) -- (B) -- (O) -- cycle;
\fill[gray2] (B) -- (C) -- (O) -- cycle;
\fill[gray3] (C) -- (A) -- (O) -- cycle;

\draw[thick] (A) -- (B) -- (C) -- cycle;

\draw[thin] (A) -- (O);
\draw[thin] (B) -- (O);
\draw[red,line width=0.8] (C) -- (O);

\node at (0.5,0.259-0.03) {\color{orangePython}$c=2$};

\node at (0.25,0.135) {$c=\frac{8}{5}$};
\node at (0.75,0.135) {$c=\frac{8}{5}$};
\node[red] at (0.5,0.64)   {$c=\frac{8}{5}$};

\node at (0.33,0.36) {SPT$_{\omega}$};
\node at (0.67,0.36) {SPT$_{\overline{\omega}}$};
\node at (0.5,0.07)  {Trivial};

\node[draw, rectangle, minimum size=0.02cm, fill=bluePython]  at (0.5,{sqrt(3/4)/(2-0.9) }) {};
\node[draw, rectangle, minimum size=0.02cm, fill=tealPython]  at (0.5,{sqrt(3/4)/(2+0) }) {};
\node[draw, rectangle, minimum size=0.02cm, fill=greenPython] at (0.5,{sqrt(3/4)/(2+0.9) + 0.005}) {};

\draw[fill=orangePython] (0.5,{sqrt(3/4)/(2+1) - 0.005}) circle (0.02cm);

\node at (0.5,-0.06) {
    \begin{tikzpicture}[scale=0.65]
    \foreach \x in {0,1.5,3}{
    \filldraw[blue,opacity=0.15] (0.25+\x,0) ellipse (12pt and 5pt);
    \filldraw[blue,opacity=0.15] (0.25+0.5+\x,0) ellipse (12pt and 5pt);
    };
    \foreach \x in {0,1,2,3,4,5,6,7,8}{
    \filldraw (0.5*\x,0) circle (1.5pt);
    };
    \end{tikzpicture}
};

\node[rotate=60] at (0.25-0.05,0.866/2+0.03) {
    \begin{tikzpicture}[scale=0.65]
    \foreach \x in {0.5,2}{
    \filldraw[blue,opacity=0.15] (0.25+\x,0) ellipse (12pt and 5pt);
    \filldraw[blue,opacity=0.15] (0.25+0.5+\x,0) ellipse (12pt and 5pt);
    };
    \filldraw[blue,opacity=0.15] (0.25+3.5,0) ellipse (12pt and 5pt);
    \foreach \x in {0,1,2,3,4,5,6,7,8}{
    \filldraw (0.5*\x,0) circle (1.5pt);
    };
    \end{tikzpicture}
};

\node[rotate=-60] at (0.75+0.05,0.866/2+0.03) {
    \begin{tikzpicture}[scale=0.65]
    \foreach \x in {0.5,2}{
    \filldraw[blue,opacity=0.15] (0.25+\x,0) ellipse (12pt and 5pt);
    \filldraw[blue,opacity=0.15] (0.25+0.5+\x,0) ellipse (12pt and 5pt);
    };
    \filldraw[blue,opacity=0.15] (0.25+3.5,0) ellipse (12pt and 5pt);
    \foreach \x in {0,1,2,3,4,5,6,7,8}{
    \filldraw (0.5*\x,0) circle (1.5pt);
    };
    \end{tikzpicture}
};

\end{tikzpicture}
};
\node at (7.5,-0.3){
    \includegraphics[scale=0.5]{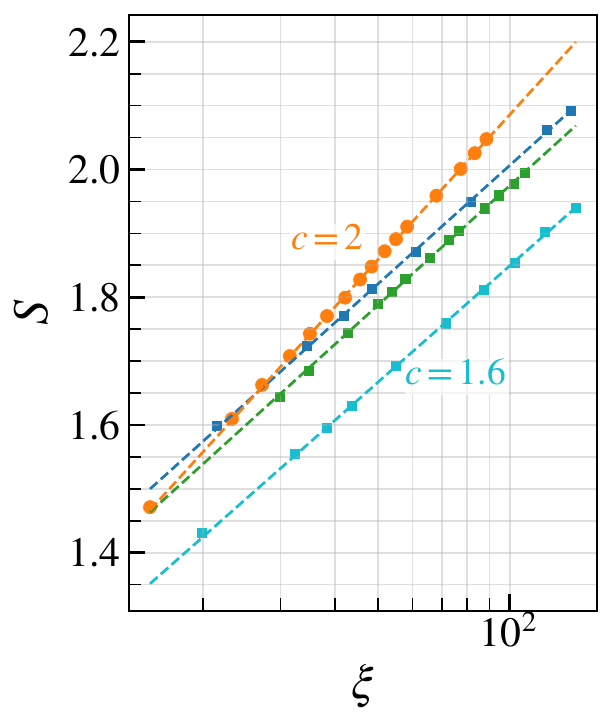}
};
\node at (-3.5,2.8) {(a)};
\node at (4.5,2.8) {(b)};
\end{tikzpicture}

\caption{\textbf{Gapped and gapless SPTs and multicriticality in the trimerized qutrit XX chain.} (a) The phase diagram for $H(J=0, a_0, a_1, a_2)$ in Eq.~\eqref{eq:XXYY}, which is nearest-neighbor chain of qutrits with a three-site unit cell. Three gapped SPT phases protected by the unitary $\mathbb Z_3^X \times \mathbb Z_3^Z$ symmetry are separated by three purely gapless SPT phases protected by the same symmetry group with central charge $c=\frac{8}{5}$. The red line is a $c=\frac{8}{5}$ CFT which has stable edge modes despite being critical. These three lines meet at a multicritical point with an LSM anomaly and central charge $c=2$. (b) We estimate the central charge of the SPT transitions by calculating entanglement entropy and correlation lengths at finite increasing bond dimensions in DMRG, and fitting $S = \frac{c}{6}\log(\xi)+\textrm{const.}$ at $H = H_0 + \lambda(H_1 + H_2) $ for $\lambda = 1$ (orange circle) and $\lambda =  0.95, 0.5, 0.05$ (square).
}
\label{fig:triange_diagram}
\end{figure}

Guided by these considerations, we have arrived at a concrete family of models: the (trimerized) qutrit XXYY chain, defined by the Hamiltonian
\begin{equation}
\boxed{ H(J,a) \equiv \sum_j a_{j \bmod 3} \left( X_j^\dagger X_{j+1} +Z_j^\dagger Z_{j+1} + J( Y^\dagger_j Y_{j+1} + W_{j}^\dagger W_{j+1}) +  {\rm h.c.} \right) \equiv a_0 H_0 + a_1 H_1 + a_2 H_2 } \; .
\label{eq:XXYY}
\end{equation}
The coefficients $a_{0}, a_1, a_2 \geq 0$ introduce a spatial modulation with period three, which we also refer to as trimerization. This modulation explicitly breaks the $\mathbb{Z}_3$ entangler symmetry while preserving the protecting $\mathbb{Z}_3 \times \mathbb{Z}_3$ on-site symmetry, and is the key ingredient that enables flows into distinct gapped \emph{and} gapless SPT phases. The phase diagram for different $a_j$ is illustrated in Figure \ref{fig:triange_diagram}, at which we arrive over the course of this work.

We refer to the case $J=0$ as the qutrit XX model, previously introduced as the ``quantum torus'' model in the translation-invariant case $a_0=a_1=a_2$ \cite{Qin_2012}. Subsequent work \cite{Alavirad_2021} established numerically that for all $J\ge 0$, the translation-invariant XXYY chain realizes conformal field theories with central charge $c=2$, and also emphasized that this model provides a natural setting for the Lieb–Schultz–Mattis anomaly. The point $J=1$ is the integrable Uimin–Lai–Sutherland Hamiltonian, described by the $SU(3)_1$ Wess–Zumino–Witten theory \cite{Uimin1970, ULS_Lai, ULS_Sutherland, Itoi_1997}. In particular, we do not consider the well-known $J=1$ as our candidate model, because the enhanced $SU(3)$ symmetry forbids CFTs with $c<2$ and thus would forbid any renormalization group flows into gapless SPTs. We are not aware of any study of the model for $J<0$, which is beyond the scope of the present work. Our interest is particularly on the effect of trimerization, which to the best of our knowledge has not been studied before.

We will explore the physics of this model in increasing order of complexity: first the gapped SPT phases protected by $\mathbb Z_3 \times \mathbb Z_3$ symmetry, then their phase transitions which give distinct gapless SPT phases protected by the same symmetry group, and then finally we approach the multicriticality between these distinct gapless SPTs.

\section{Gapped SPT Phases}

Much like the $\mathbb{Z}_2 \times \mathbb{Z}_2$ case reviewed earlier, translation-breaking perturbations in our qutrit model drive RG flows into distinct gapped SPT phases. The structure is formally analogous, but the $\mathbb{Z}_3 \times \mathbb{Z}_3$ symmetry hosts \textit{three} different gapped SPT phases: SPT$_{1}$ (trivial), SPT$_\omega$ and SPT$_{\overline\omega}$. We arrive at ternary phase diagram in Figure \ref{fig:triange_diagram}(a), in which all the gapped $\mathbb{Z}_3 \times \mathbb{Z}_3$ SPT phases are realized.

\subsection{Fixed-point limits and projective edge modes}

\begin{figure}[t]
\begin{tikzpicture}
\tikzset{
triplet/.style={draw=red, fill=red, fill opacity=0.35, line width=0pt},
special/.style={draw=teal!70!black, fill=teal!40, rounded corners=3pt, line width=0.5pt, opacity=0.7},
blockfill/.style={black, opacity=0.09}
}

\foreach \y in {0,1,2}{
\foreach \x in {1,4,7,10}{
\filldraw[blockfill] (\x-1.33,-\y-0.35) rectangle (\x+1.33,-\y+0.35);
};
\foreach \x in {0,1,2,3,4,5,6,7,8,9,10,11}{
\filldraw[black] (\x,-\y) circle (2pt);
};
}

\node at (-1.5,0)  {SPT$_1$:};
\node at (-1.5,-1) {SPT$_{\omega}$:};
\node at (-1.5,-2) {SPT$_{\overline{\omega}}$:};

\foreach \x in {0,3,6,9}{
\filldraw[triplet] (\x+1,0) ellipse (1.3 and 0.3); 
}
\foreach \x in {1,4,7}{
\filldraw[triplet] (\x+1,-1) ellipse (1.3 and 0.3); 
}
\foreach \x in {2,5,8}{
\filldraw[triplet] (\x+1,-2) ellipse (1.3 and 0.3); 
}

\filldraw[special] (10-0.55,-1-0.22) rectangle (11+0.55,-1+0.22);
\filldraw[special] (0-0.55,-2-0.22) rectangle (1+0.55,-2+0.22);

\filldraw[special] (-0.25,-1-0.25) rectangle (0.25,-1+0.25);
\filldraw[special] (11-0.25,-2-0.25) rectangle (11+0.25,-2+0.25);

\node[blue] at (0,-1.3)   {$\boldsymbol{\omega}$};
\node[blue] at (10.5,-1)  {$\boldsymbol{\overline \omega}$};
\node[blue] at (11,-2.3)  {$\boldsymbol{\omega}$};
\node[blue] at (0.5,-2)   {$\boldsymbol{\overline \omega}$};
\end{tikzpicture}
\caption{\textbf{Exact fixed-point wavefunctions for the three gapped SPT phases of the trimerized qutrit XX chain.} Each red ellipse represents the totally antisymmetric $SU(3)$ singlet $\epsilon^{\mu\nu\rho} \ket{\mu}\ket{\nu}\ket{\rho}$.
We observe that for our convention of three-site unit cells, ${\rm SPT}_1$ has no edge modes, whereas ${\rm SPT}_\omega$ and ${\rm SPT}_{\overline \omega}$ carry distinct projective representations of $\mathbb Z_3 \times \mathbb Z_3$ on the edge.
The same ground states apply for the trimerized qutrit XXYY chain for $J \geq 0$. }
\label{fig:qutrit_fixed_points}
\end{figure}
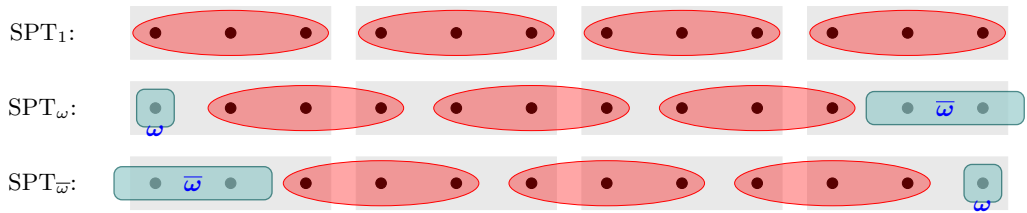

Analogous to the qubit model, the fixed-point limit is when either $a_1$, $a_2$ or $a_0$ is zero. Without loss of generality, we can call $a_0 = 0$ the trivial phase. This corresponds to a convention with unit cells on three-site blocks $(3j+1,3j+2,3j+3)$. Using the notation of Eq.~\eqref{eq:XXYY}, this is $H = a_1 H_1 + a_2 H_2$ for $a_1,a_2>0$. The ground state turns out to be independent of the precise value of $a_1,a_2>0$ and $J \geq 0$, and it is given by the totally-antisymmetric $SU(3)$ singlet on every unit cell, as shown in Figure \ref{fig:qutrit_fixed_points}. Translation by one site acts as the entangler between the three SPT phases.

As in the qubit case, the distinction between SPT phases becomes manifest on a finite chain from site $1$ to $3N$. For $a_1 = 0$ or $a_2 = 0$, the ground states exhibit degenerate edge modes corresponding to decoupled qutrits as shown in Figure \ref{fig:qutrit_fixed_points}. These edge qutrits transform under distinct nontrivial projective representations of $\mathbb{Z}_3^X \times \mathbb{Z}_3^Z$, confirming that the two phases are symmetry-distinct. This gives us all three distinct SPT phases protected by $\mathbb Z_3 \times \mathbb Z_3$ symmetry ~\cite{Duivenvoorden13a,Duivenvoorden13b,geraedts2014exactmodelssymmetryprotectedtopological,Santos15}.

\subsection{String order parameters}

Like in the qubit case, we can define string order parameters

\begin{equation}
\begin{split}
    S^{(0)}_j &= \left(\prod_{k<j} X_{3k+1}X_{3k+2}X_{3k+3} \right)\\
    S^{(1)}_j &= \left(\prod_{k<j} X_{3k+1}X_{3k+2}X_{3k+3}\right) X_{3j+1}\\
    S^{(2)}_j &= \left(\prod_{k<j} X_{3k+1}X_{3k+2}X_{3k+3}\right) X_{3j+1} X_{3j+2}
\end{split}
\label{eq:string_order_params}
\end{equation}

Assuming the unit-cell blocks $(3j+1, 3j+2, 3j+3)$, these string order parameters carry distinct charges $1, \omega, \omega^2$ respectively under the $\mathbb{Z}_3^Z$ subgroup generated by $\prod_j Z_j$. For the fixed point ground states for trivial, SPT-1, and SPT-2, it is clear that only $S^{(0)}_j$, $S^{(1)}_j$, and $S^{(2)}_j$ respectively have long-range order $\lim_{|j_L - j_R| \to \infty} \langle S^{(\alpha)\dagger}_{j_1}S^{(\alpha)}_{j_2} \rangle$.

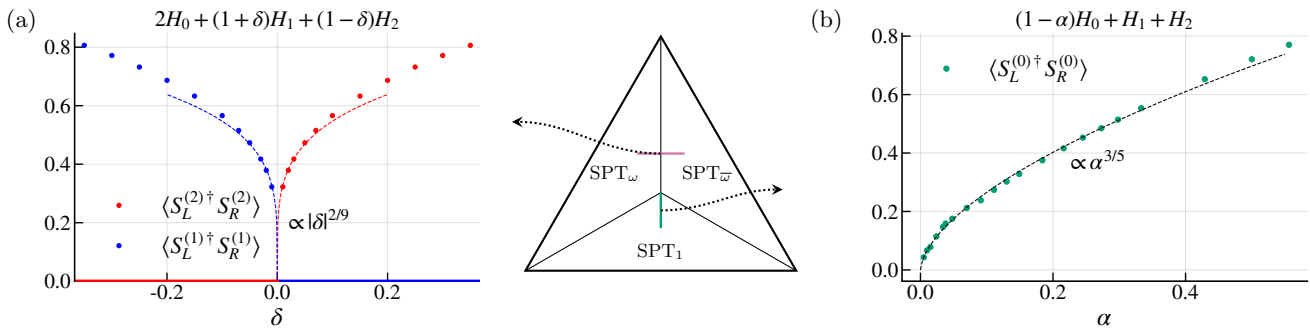
\begin{figure}
    \centering
\begin{tikzpicture}
\node at (-3.7,2) {(a)};
\node at (7,2) {(b)};
\node at (-0.5,0){
\includegraphics[scale=0.4]{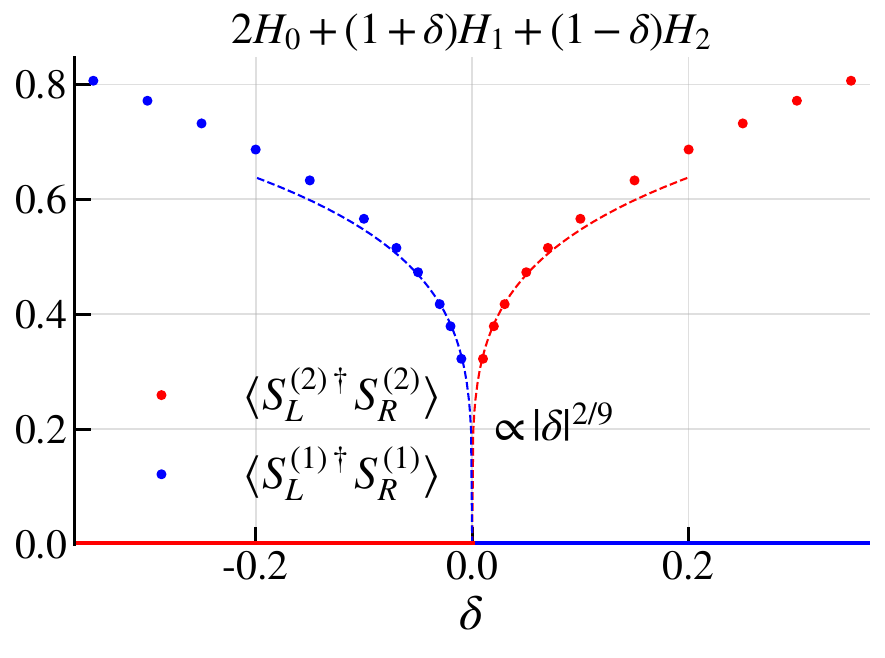}
};
\node at (4.8,0.3){
    \begin{tikzpicture}[scale=3.6, every node/.style={scale=0.8}]
        \path[use as bounding box] (-0.05,-0.05) rectangle (1.05,0.95);

          \definecolor{deltaColor}{HTML}{CC79A7}   
          \definecolor{alphaColor}{HTML}{009E73}   
          
          \coordinate (A) at (0,0);
          \coordinate (B) at (1,0);
          \coordinate (C) at (0.5,0.866);

          \draw[thick] (A) -- (B) -- (C) -- cycle;
          \coordinate (O) at (0.5,0.2886667);

          \draw (A) -- (O);
          \draw (B) -- (O);
          \draw (C) -- (O);

          \coordinate (S1L) at (0.4125,0.4330);
          \coordinate (S1R) at (0.5875,0.4330);
          \draw[line width=0.9pt, color=deltaColor] (S1L) -- (S1R);

          \coordinate (S2a) at (0.5,0.2886667);
          \coordinate (S2b) at (0.5,0.157459164);
          \draw[line width=0.9pt, color=alphaColor] (S2a) -- (S2b);

          \node at (0.33,0.36) {SPT$_{\omega}$};
          \node at (0.67,0.36) {SPT$_{\overline{\omega}}$};
          \node at (0.50,0.07)  {SPT$_1$};

            \coordinate (S1mid) at ($(S1L)!0.5!(S1R)$);
            \draw[overlay, densely dotted, line width=0.8pt,->,>=stealth]
                  (S1mid) to[out=180,in=0] (-0.05,0.55); 
            
            \coordinate (S2mid) at ($(S2a)!0.5!(S2b)$);
            \draw[overlay, densely dotted, line width=0.8pt,->,>=stealth]
                  (S2mid) to[out=0,in=180] (0.95,0.30); 

        \end{tikzpicture}
};
\node at (10.5,0){
\includegraphics[scale=0.4]{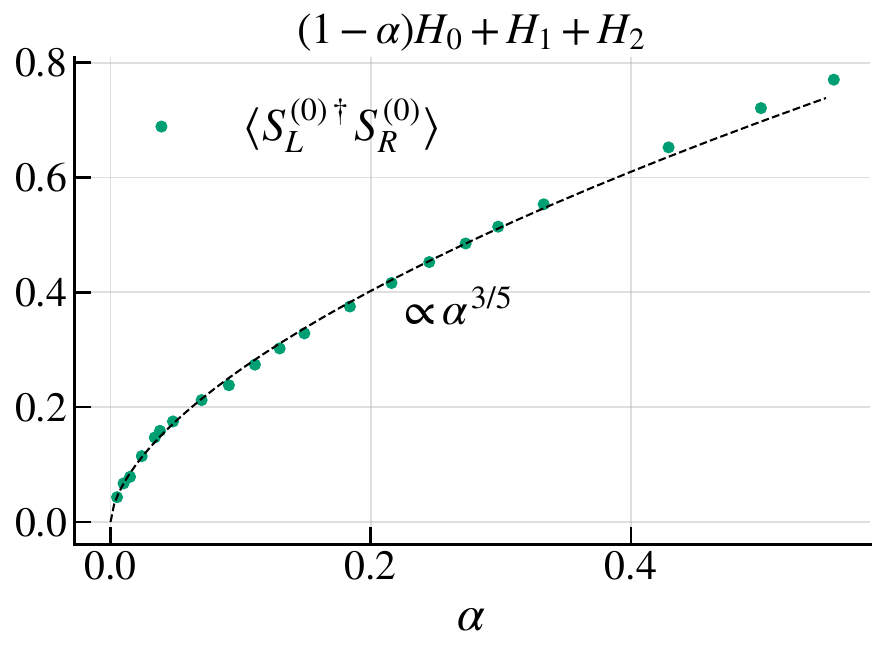}
};
\end{tikzpicture}
    \caption{\textbf{Asymptotic values of string order parameters for the SPT$_\omega$, SPT$_{\overline{\omega}}$, and SPT$_1$ phases.} We use iDMRG to obtain the string order parameters associated to the $\mathbb Z_3^X$ symmetry string (see Eq.~\eqref{eq:string_order_params}) in the trimerized qutrit XX chain (Eq.~\eqref{eq:XXYY} with $J=0$). For each gapped SPT phase, one and only one of the three string order parameters in Eq.~\ref{eq:string_order_params} has nonzero long range order. (a) Around the gapless SPT (gSPT$_{\omega,\overline \omega}$ shown here), the long-range order vanishes as $|\delta|^{2/9}$, where $a_0 = 2$ and $a_{1,2} = 1\pm \delta$. (b) In the vicinity of the multicritical point with $a_0 = 1-\alpha$ and $a_{1,2} = 1$, it vanishes as $\alpha^{3/5}$.
    \label{fig:sptlro}}
\end{figure}

Numerical calculations indicate that indeed these long range orders persist throughout what is labeled as the respective SPT phase in Fig.~\ref{fig:triange_diagram}. 
We performed ground state simulations using the infinite density matrix renormalization group (iDRMG) method \cite{White_1992, White_1993, Hauschild_2018}, using bond dimension $\chi \lesssim 600$. From the leading eigenvectors of the transfer matrices, we can compute the long-range order (see Appendix \ref{app:numerics} for the procedure). The results are shown in Figure \ref{fig:sptlro}. In each SPT phase, only the respective string order parameter has a nonzero long range order; the other two vanish. Refer to Sections \ref{subsec:why-orbifold-potts2} and \ref{subsec:multicrit} for discussion about critical exponents that appear in the figure.

\section{Gapless SPTs and Multicriticality \label{sec:gapless}}

Breaking translational symmetry can give rise not only to gapped phases, but also to \textit{gapless} ones. These gapless phases appear precisely at continuous phase transitions between distinct gapped SPTs, marking a key difference from the qubit analog and highlighting the richer structure of the qutrit case. Numerically, such gapless behavior is observed along the $J=0$ slice (the qutrit XX model); we will return later to explain why this slice is special. 

\subsection{Central charge from tensor network simulations \label{subsec:central}}

While we present an analytic approach in Section~\ref{subsec:analytic}, we first discuss numerical results. 
For the gapless SPTs we also performed iDMRG with bond dimension $\chi \lesssim 600$.
For each fixed choice of bond dimension, a critical system will converge to have an effective bipartite entanglement entropy $S(\chi)$ and correlation length $\xi(\chi)$, both of which diverge as $\chi \to \infty$. Fitting $S(\chi) = \tfrac{c}{6}\ln(\xi(\chi)) + \text{const.}$, as outlined in Ref.~\onlinecite{Pollmann_2009}, reveals three critical rays in the phase diagram of Figure \ref{fig:triange_diagram}, each with central charge $c=8/5$. These three rays corresponds to distinct gapless SPTs, as we discuss in Section \ref{subsec:gSPT_SOP}. Translation, as the entangler, interchanges them. In Section \ref{subsec:analytic}, we show that the SPT$_\omega$ $\leftrightarrow$ SPT$_{\overline{\omega}}$ transition here coincides with the $\mathbb{Z}_3 \times \mathbb{Z}_3$ cluster SPT transition studied by the authors in Ref.~\onlinecite{Prembabu_2022} in collaboration with Ryan Thorngren. We note that these gapless SPT theories are not anomalous; the anomalous symmetry exists only at the multicritical point (Sec.~\ref{subsec:multicrit}) and is explicitly broken along the critical lines.

\subsection{Identifying distinct gapless SPTs via quasi-long-range string order \label{subsec:gSPT_SOP}}

In gapped SPT phases, a standard bulk invariant is given by the charge of the lowest-energy state in the twisted sector, which is equivalent to the charge of the nonvanishing string order parameter. An analogous structure holds for gapless SPT phases \cite{Verresen_2021}. In particular, gapless SPTs related by an entangler can be distinguished by the ground-state charges of their twisted sectors, which thus play the role of topological invariants. By state–operator correspondence, these ground states map to the string operators of lowest scaling dimension, termed ``symmetry fluxes''. At critical points between gapped SPTs, the relevant symmetry fluxes are naturally inherited from the neighboring gapped phases. 

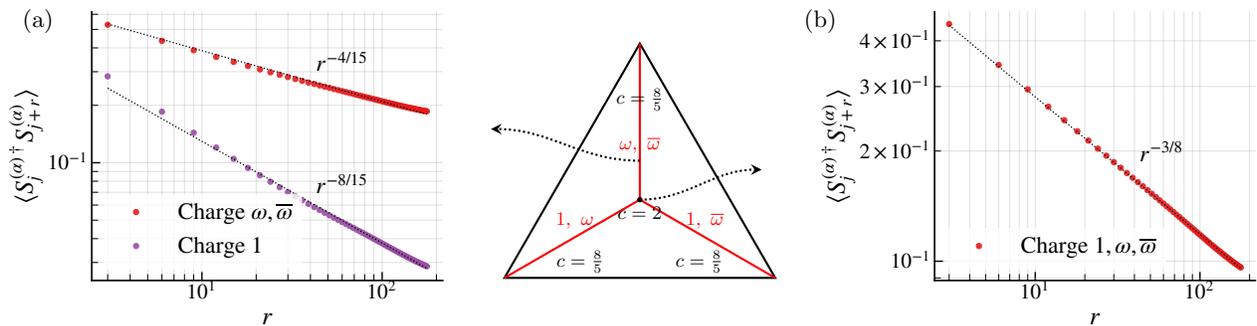
\begin{figure}
\begin{tikzpicture}
\node at (-3,2){(a)};
\node at (7.4,2){(b)};
\node at (-0.5,0){
\includegraphics[scale=0.4]{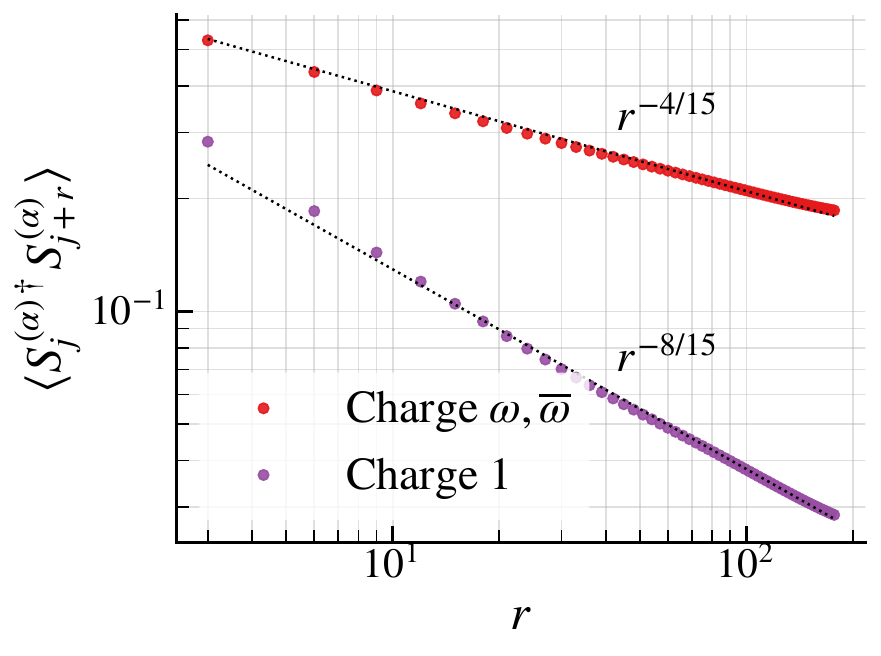}
};
\node at (5,0.2){
    \begin{tikzpicture}[scale=3.6, every node/.style={scale=0.8}]
            \path[use as bounding box] (-0.05,-0.05) rectangle (1.05,0.95);

            \coordinate (A) at (0,0);
            \coordinate (B) at (1,0);
            \coordinate (C) at (0.5,0.866);
            \coordinate (Center) at (0.5,0.289); 
            \coordinate (RefPoint) at (0.5,0.433);

            \draw[thick] (A) -- (B) -- (C) -- cycle;
            \draw[red, thick] (A) -- (Center);
            \draw[red, thick] (B) -- (Center);
            \draw[red, thick] (C) -- (Center);
            \fill[black] (Center) circle (0.01);

            \draw[overlay, densely dotted, line width=0.8pt,->,>=stealth]
                (RefPoint) to[out=180,in=0] (-0.05,0.55);
    
            \draw[overlay, densely dotted, line width=0.8pt,->,>=stealth]
                (Center) to[out=0,in=180] (0.95,0.4);
    
            \node at (0.5,0.24) {$c=2$};
            \node at (0.28,0.06) {$c=\tfrac{8}{5}$};
            \node at (0.72,0.06) { $c=\tfrac{8}{5}$};
            \node at (0.5,0.66) { $c=\tfrac{8}{5}$};

            \node at (0.26,0.21) {\color{red} $1,\ \omega$};
            \node at (0.74,0.21) {\color{red}  $1,\ \overline{\omega}$};
            \node at (0.5,0.49) {\color{red}$\omega,\ \overline{\omega}$};
        \end{tikzpicture}
};
\node at (10.3,0){
\includegraphics[scale=0.4]{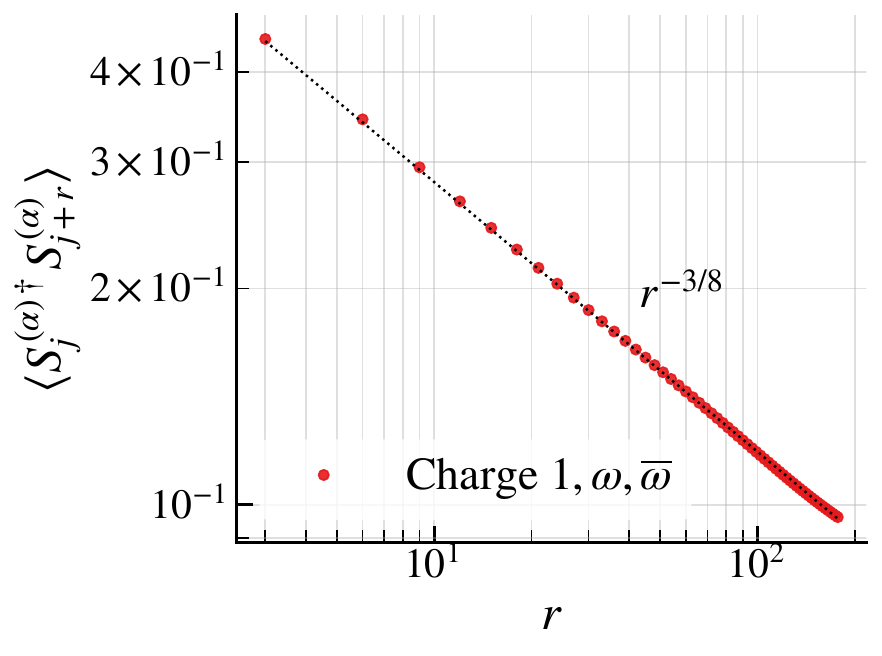}
};
\end{tikzpicture}
    \caption{
    \textbf{Detecting gapless SPTs with quasi-long-range string order.}
    Here we plot  $\mathbb Z_3^X$ string two-point functions $\langle S^{(\alpha)\dagger}_{j} S^{(\alpha)}_{j+r}\rangle$ for $\alpha = 0,1,2$ in the gapless regimes. The string operator $S^{(\alpha)}$ defined in Eq.~\eqref{eq:string_order_params} has charge $\omega^\alpha$ under the other symmetry subgroup $\mathbb Z_3^Z $. (a) For each gapless SPT, a bulk topological invariant is given by the charges of the lightest twisted-sector fields. For instance in the gSPT$_{\omega, \overline\omega}$ at $H_0 + 0.5 H_1 + 0.5 H_2$, the $\omega$ and $\overline\omega$ charged strings have the slowest-decaying correlations (red), while the neutral string decays faster (purple). Inequivalent gapless SPTs have different charges, providing a topological invariant distinguishing the three gapless SPTs. The scaling exponents are consistent with the orbifold Potts$^2$ CFT. (b) At the $c=2$ multicritical point the differently-charged string order parameters all have identical dimension $\approx 3/16$ due to being related by the entangler symmetry, indicating the mixed anomaly.
    }
    \label{fig:string_charge_gapless_SPT}
\end{figure}

To analyze the gapless SPTs, we therefore examine their string order parameters, given in Eq.~\ref{eq:string_order_params}. For each critical $c=8/5$ ray in Figure \ref{fig:triange_diagram}, we calculate the two-point function $\langle S^{(\alpha)\dagger}_{j_1}S^{(\alpha)}_{j_2} \rangle$ (for the $\mathbb{Z}_3^X$ symmetry string) and study its asymptotic behavior $\sim 1/|j_1-j_2|^{2\Delta}$. We find that all three string correlation functions decay algebraically, consistent with the SPT being \emph{purely} gapless (i.e., no gapped symmetry sectors). However, they do not all have the same scaling dimension: as shown in Figure \ref{fig:string_charge_gapless_SPT} two string correlation functions decay with an exponent $\frac{4}{15}$ (implying a scaling dimension $\Delta = \frac{2}{15}$) whereas the third decays with a doubled exponent, $\frac{8}{15}$.

Interestingly, the $\mathbb{Z}_3^Z$ charges of the two slowest-decaying correlation depends on which $c=\frac{8}{5}$ line we are on. 
For instance, along the SPT$_\omega$ $\leftrightarrow$ SPT$_{\overline\omega}$ transition, the lightest strings carry charges $\omega$ and $\overline{\omega}$, and translation by one site cyclically permutes them by multiplying the charge by $\omega$. 
This establishes the sharp distinction between the three gapless phases. Since the scaling dimensions $\frac{4}{15}$ and $\frac{8}{15}$ are discretely distinct, there is no way of altering the charge assignments whilst remaining in the low-energy CFT. (Moreover, as will see, this CFT has no marginal parameters, and hence these scaling dimensions are robust.) In conclusion, as long as we preserve the $\mathbb Z_3^X \times \mathbb Z_3^Z$ symmetry, these are distinct symmetry-enriched CFTs.
Based on these twisted sector charges, we denote the gSPT between SPT$_\omega$ and SPT$_{\overline \omega}$ as gSPT$_{\omega, \overline{\omega}}$, and likewise the other two gSPTs as gSPT$_{1, \omega}$ and gSPT$_{1,\overline{\omega}}$.

\subsection{Why orbifold Potts$^2$ CFT?
\label{subsec:why-orbifold-potts2}}

In passing we have already alluded to the fact that the $c=\frac{8}{5}$ lines are described by an orbifold of the Potts$^{2}$ CFT. While we give a derivation of this fact (in a limit of our model) in Section~\ref{subsec:analytic}, let us briefly describe what this CFT is and why it is natural to expect it as a phase transition between these gapped SPT phases.

The Potts CFT is one of the celebrated minimal unitary CFTs \cite{Ginsparg}. It is the only such CFT with a $\mathbb Z_3$ symmetry, and it describes a transition between a $\mathbb{Z}_3$-SSB phase (where the spin field $\sigma$ with scaling dimension $\Delta = \frac{2}{15}$ condenses) and a nondegenerate paramagnetic gapped phase (where the dual disorder field $\mu$ condenses, also with $\Delta = \frac{2}{15}$). Since this CFT has a central charge $c=\frac{4}{5}$, it might naively seem suggestive that our system with a $\mathbb Z_3 \times \mathbb Z_3$ symmetry and central charge $c=\frac{8}{5} = 2 \times \frac{4}{5}$ could be described by a `double copy' of the Potts CFT. However, this cannot quite work, since such a `Potts$^2$' CFT can only describe transitions between symmetry-breaking phases of matter.

This issue can be fixed by a simple trick called orbifolding \cite{Ginsparg}, or equivalently discrete gauging, which means the nearby phase diagram of the CFT can now contain distinct SPT phases as first pointed out by Ref.~\onlinecite{Tsui_2017}. To see how this plays out, let us start with a double copy of the Potts CFT, where we label the local spin fields as $\sigma_A$ and $\sigma_B$ and the nonlocal disorder operators as $\mu_A$ and $\mu_B$. This has a $\mathbb Z_3 \times \mathbb Z_3$ symmetry under which $\sigma_A$ and $\sigma_B$ are charged. The act of orbifolding entails that we gauge the `diagonal' $\mathbb Z_3$ subgroup under which $\sigma_A$ and $\sigma_B$ carry the same charge. This has two major consequences: (i) the fields $\sigma_A$ and $\sigma_B$ are now no longer well-defined and thus `projected out', although $\sigma_A \sigma_B^\dagger$ is gauge-neutral and thus remains a valid local operator of our orbifolded theory, and (ii) although $\mu_A$ and $\mu_B$ are nonlocal operators, their product $\mu_A \mu_B$ is now a local scaling operator in the orbifolded theory, since their `string' is made up out of the diagonal $\mathbb Z_3$ symmetry which is now invisible (since it has been gauged!).

\begin{table}
\centering
\begin{tabular}{|c|c|c|}
\hline
 & \textbf{Potts$^2$} & \textbf{Orbifold Potts$^2$ (diag.\ $\mathbb{Z}_3$)} \\
\hline
\textbf{Local operators} 
& $\sigma_A,\, \sigma_B,\, \sigma_A\sigma_B ,\;\ldots$ 
& $\sigma_A\sigma_B^\dagger,\; \mu_A\mu_B,\;\ldots$ \\
\hline
\textbf{Nonlocal operators} 
& $\mu_A,\, \mu_B,\mu_A\mu_B ,\;\ldots$ 
& $\sigma_A,\, \sigma_B,\, \mu_A,\, \mu_B,\;\ldots$ \\
\hline
\end{tabular}
\caption{Local vs.\ nonlocal operators before and after orbifolding the diagonal $\mathbb{Z}_3$. \label{table:orbifold}}
\end{table}

In summary, orbifolding can be seen as a reshuffling of which scaling operators are regarded as local and which as nonlocal (see Table~\ref{table:orbifold} for a summary). While this does not affect certain properties like the central charge, it can have dramatic effects on how to interpret the nearby phase diagram \cite{Tsui_2017}. To see this, let us name our local fields of the orbifold theory as $\sigma \equiv \sigma_A \sigma_B^\dagger$ and $\tilde \sigma \equiv \mu_A \mu_B$, and our nonlocal fields as $\mu \equiv \mu_A$ and $\tilde \mu \equiv \sigma_B$. Note that with this choice, most operators commute with one another, except for the pair $\sigma$ and $\mu$, as well as the pair $\tilde \sigma$ and $\tilde \mu$, which satisfy the usual commutation relations for dual `order/disorder' pairs. We thus see that this orbifold theory still has $\mathbb Z_3 \times \mathbb Z_3$ symmetry, with $\sigma$ and $\tilde \sigma$ as order operators and $\mu$ and $\tilde \mu$ as disorder operators.

We can explain the nearby phase diagram in terms of these new operators. For instance, one of the phases proximate to the original Potts$^2$ theory has both $\sigma_A$ and $\mu_B$ condensed (i.e., $\mathbb Z_3^A$ is spontaneousy broken whereas $\mathbb Z_3^B$ is preserved in the paramagnetic phase). Since $\sigma_A = \sigma \tilde \mu$ and $\mu_B = \mu^\dagger \tilde \sigma $, we learn that in the orbifold theory this phase has non-trivial string order parameters, where, e.g., the $\tilde \mu$ domain walls come with $\sigma$ charge. This is thus a non-trivial SPT phase! One can similarly check that the phase where $\mu_A$ and $\sigma_B$ are condensed maps to the trivial phase in the orbifold theory. Ref.~\onlinecite{Tsui_2017} pointed out that this makes this CFT a natural candidate for describing $\mathbb Z_3 \times \mathbb Z_3$ SPT transitions, and Ref.~\onlinecite{Prembabu_2022} pointed out that there are multiple distinct symmetry-enriched versions of this CFT.

We note that this orbifold Potts$^2$ CFT also naturally clarifies the exponents we have observed thus far. In particular, the two dominant disorder scaling operators are $\mu$ and $\tilde \mu$, whose scaling dimensions are those of $\mu_A$ and $\sigma_B$ of the original Potts$^2$ CFT, i.e., $\Delta = \frac{2}{15}$, whereas the field $\mu \tilde \mu$ has dimension $\Delta = \frac{4}{15}$, consistent with Figure \ref{fig:string_charge_gapless_SPT}. Moreover, the critical exponent $\frac{2}{9}$ of the vanishing of gapped SPT string order parameters in Figure \ref{fig:sptlro} also agrees with the orbifold Potts$^2$ CFT. It is equal to $\frac{2\Delta_{\mu}}{1-\Delta_{\epsilon}}$ where $\Delta_{\epsilon}$ is the dimension of the Potts thermal operator that perturbs into the SPT phases. 

In the following section we will derive this CFT analytically.

\subsection{Analytic derivation of orbifold Potts$^2$ CFT \label{subsec:analytic}}

We now give an analytic derivation of one of the $c=8/5$ critical theories 
in the qutrit XX model. 
We show that near the corner limit $a_0 \gg a_1,a_2$, a projection onto the low-energy subspace reduces the model to the $\mathbb{Z}_3 \times \mathbb{Z}_3$ 
cluster SPT chain. 
Since the critical point of the cluster chain realizes the diagonal $\mathbb{Z}_3$ 
orbifold of two Potts models (``orbifold Potts$^2$'') 
\cite{Tsui_2017, geraedts2014exactmodelssymmetryprotectedtopological, Prembabu_2022}, 
this limit of the qutrit XX model provides a direct microscopic realization of the 
same $c=8/5$ CFT. 
In the following subsection we make this projection explicit. 

This projection demonstrates two complementary statements: On one hand, the qutrit XX model in this parameter regime has the same spectrum 
and scaling behavior as the orbifold Potts$^2$ theory; 
on the other hand, once symmetry charges are taken into account, it realizes distinct 
gapless SPT versions of that theory.

\subsubsection{Projection onto $H_0$ ground state \label{subsubsec:projection}}

It will be convenient to perform a local change of basis inside the three-site unit cells such that the projection on to the ground state of $H_0$ is equivalent to freezing out certain auxiliary sites (we remind the reader that $H_0$, $H_1$, $H_2$ are defined in Eq.~\eqref{eq:XXYY}). To do so we apply the following unitary map\footnote{This can be realized ($\mathcal{O} \to \mathcal{O}'$ means $U\mathcal{O}U^\dagger = \mathcal{O}'$)  through the following sequence of local gates, where $H$ refers to Hadamard ($X \to Z \to X^\dagger$) :
\begin{equation}
    U_0 = \prod_j H^{\dagger}_{3j} CZ^\dagger_{3j, 3j+1} H_{3j} H_{3j+1} CZ_{3j, 3j+1} H_{3j}
\end{equation}
} $U_0$ and index re-labeling, which sends $H_0$ to a sum of single-site terms on sites $2j+\frac{1}{2}$ which should be understood to be frozen auxiliary sites (in the large $a_0$ limit) distinct from the dynamical sites on integer indices $2j, 2j+1$:

\begin{equation}
    \begin{split}
        X_{3j} &\to Z_{2j} Z_{2j+\frac{1}{2}} \qquad Z_{3j} \to X_{2j} X_{2j+\frac{1}{2}} \\
        X_{3j+1} &\to Z_{2j} Z_{2j+\frac{1}{2}}^{\dagger} \qquad Z_{3j+1} \to X_{2j} X_{2j+\frac{1}{2}}^{\dagger} \\
        \mathcal{O}_{3j-1} &\to \mathcal{O}_{2j-1}
    \end{split}
\end{equation}

After applying $U_0$, for further technical convenience we can apply the unitary $U_1 \equiv \prod_j CZ^\dagger_{2j-1, 2j} CZ^\dagger_{2j, 2j+1}$. Together, the finite-depth local unitary $U_1 U_0$ sends

\begin{equation}
\begin{split}
H_0 \to&\ \tilde H_0 \equiv \sum_j {\color{red} Z_{2j+\frac{1}{2}}} + {\color{red} X_{2j+\frac{1}{2}}} + \text{h.c.} \\
H_1 \to&\ \tilde H_1 \equiv \sum_j Z_{2j}^\dagger {\color{red} Z_{2j+\frac{1}{2}}^{\dagger}} X_{2j+1}^\dagger Z_{2j+2} + Z_{2j-1}^\dagger X_{2j} {\color{red} X_{2j+\frac{1}{2}}^{\dagger}} Z_{2j+1} + \text{h.c.} \\
H_2 \to&\ \tilde H_2 \equiv \sum_j Z_{2j-2}^\dagger X_{2j-1} Z_{2j} {\color{red} Z_{2j+\frac{1}{2}}^{\dagger}} + Z_{2j-1}^\dagger X_{2j}^\dagger {\color{red} X_{2j+\frac{1}{2}}^{\dagger}} Z_{2j+1} + \text{h.c.}
\end{split}
\end{equation}

For convenience we have highlighted in red the sites we want to interpret as `auxiliary'. In the large $a_0$ limit, the auxiliary sites on $2j+\frac{1}{2}$ indices are frozen in state  $|g\rangle \equiv (|2\rangle - |1 \rangle)/\sqrt{2}$. The projector $P = \bigotimes _{j}\left(|g\>\<g|\right)_{2j+\frac{1}{2}}$   satisfies $P X_{2j+\frac{1}{2}} P = P Z_{2j+\frac{1}{2}} P = -\frac{1}{2}P $. If we project these gapped degrees of freedom into their low-energy state, the remaining two-site unit cells are coupled by the following Hamiltonians:
\begin{equation}
\begin{split}
    P \tilde H_1 P = -\frac{1}{2}\sum_j \left( Z_{2j}^{\dagger} X_{2j+1}^\dagger Z_{2j+2} + Z_{2j-1}^\dagger X_{2j} Z_{2j+1} + \text{h.c.} \right) \\
    P \tilde H_2 P  = -\frac{1}{2}\sum_j \left( Z_{2j}^{\dagger} X_{2j+1} Z_{2j+2} + Z_{2j-1}^\dagger X^\dagger_{2j} Z_{2j+1} + \text{h.c.} \right)
\end{split}
\end{equation}

In the frozen limit we \textit{exactly} recover the cluster SPT Hamiltonians. 
Explicitly, $P \tilde H_1 P$ and $P \tilde H_2 P$ are the SPT$_\omega$ and SPT$_{\overline\omega}$
cluster-chains~\cite{geraedts2014exactmodelssymmetryprotectedtopological}, up to an overall factor. 
The effective Hamiltonian coincides with the parametrization given in 
Eq.~(5) of Ref.~\onlinecite{Prembabu_2022}, with $s = (a_1-a_2)/(a_1+a_2)$. 
At the symmetric point $a_1=a_2$, the cluster chain is shown there to realize the 
orbifold Potts$^2$ theory, thus completing the identification. 

The mapping also makes the symmetry structure transparent. 
The two global $\mathbb{Z}_3$ generators $\prod_j X_j$ and $\prod_j Z_j$ 
of the qutrit XX model map onto the odd and even generators, respectively, 
of the cluster chain’s $\mathbb{Z}_3 \times \mathbb{Z}_3$ symmetry. 
Likewise, the three string order parameters $S^{(0)}_j, S^{(1)}_j, S^{(2)}_j$ 
map onto the cluster-chain string operators
\[
    \prod_{k \le j} X_{2k-1}, \quad
    \Big(\prod_{k \le j} X_{2k-1}\Big) Z_{2j}, \quad
    \Big(\prod_{k \le j} X_{2k-1}\Big) Z_{2j}^\dagger .
\]
As shown in Ref.~\onlinecite{Prembabu_2022}, the latter two operators with endpoint charge 
decorations furnish the lightest symmetry fluxes of the gapless SPT. The projection thus identifies both the low energy operator content and the symmetry properties with that of the orbifold Potts$^2$ gapless SPTs from Ref.~\onlinecite{Prembabu_2022}. We note that the boundary analysis of that work carries over to the present model. Hence, this particular transition will generically have a stable boundary edge mode, with a boundary 0+1d deconfined quantum critical point between the 0+1d symmetry-breaking edge modes.

\subsubsection{Stability and deviations}
\label{subsec:stability}

Having established the corner limit as the orbifold Potts$^2$ CFT with $c=8/5$, we now ask about the stability of this theory as we move away from the corner of the ternary phase diagram in Fig \ref{fig:triange_diagram}. 
As shown in Ref.~\onlinecite{Prembabu_2022}, when imposing time-reversal symmetry, Hadamard symmetry, and parity inversion ($a_1=a_2$), the only symmetric relevant perturbation is the scaling operator $\epsilon^A \epsilon^B$ (using the un-orbifolded Potts$^2$ notation from Appendix \ref{app:SW}) of dimension $8/5$. 
Thus, the key analytic question is whether such a perturbation is generated in Schrieffer–Wolff perturbation theory.

We analyze the regime $0<\lambda = a_1=a_2<a_0 =1$, with details given in Appendix \ref{app:SW}. Each Schrieffer–Wolff term can be matched to a symmetry-allowed local field in the CFT.
The $\mathcal{O}(\lambda)$ correction is simply the unperturbed cluster chain itself, as we derived in Section~\ref{subsubsec:projection} above.

In the Potts$^2$ notation, we find that the $\mathcal{O}(\lambda^2)$ order terms are
\begin{equation}
\Phi^{A}_{X\bar{\epsilon}} \Phi^B_{\epsilon \bar{X}}
+ \Phi^{B}_{X\bar{\epsilon}} \Phi^A_{\epsilon \bar{X}}
+ \textrm{const.}\cdot \partial_x (\epsilon^A - \epsilon^B) \, ,
\end{equation}
where $\epsilon$ is the Potts thermal operator of dimension $4/5$, and $\Phi_{X\bar \epsilon}$ and $\Phi_{\epsilon \bar X}$ are its $W$-algebra descendants of dimension $9/5$.  
Neither contribution changes the IR theory: the first is RG-irrelevant, while the second is a total derivative.  This is because the ground state and single-site excitations of on-site $H_0$ terms are odd and even respectively under charge conjugation, forcing any second order terms to be a product of charge-conjugation-odd Potts operators (on either the same chain or different chains) and thereby ruling out $\epsilon^A \epsilon^B$. 

Only at third order does the symmetric relevant operator $\epsilon^A \epsilon^B$ appear, with a positive coefficient that would normally destabilize the $c=8/5$ theory.
In the qutrit XXYY model of Eq.~\ref{eq:XXYY}, however, the additional parameter $J$ provides a stronger tuning knob: it introduces $\epsilon^A \epsilon^B$ already at first order.
Adjusting $J$ against the $a_1,a_2$ contribution can cancel the perturbation, thereby stabilizing the critical line slightly away from $J=0$. Thus this analytic picture suggests that the true orbifold Potts$^2$ gapless SPT line is shifted to small $J$. Since $\epsilon^A \epsilon^B$ is only generated at third-order in perturbation theory, it is natural to expect that $J$ should remain small, i.e., $J \approx 0$.

\subsection{Persistence of $c=8/5$}

Numerics nevertheless indicate remarkable stability of the orbifold Potts$^2$ CFT without having to tune $J$ away from 0.
DMRG finds that the entire segment of the qutrit XX model phase diagram from the $a_0$-corner of the triangle to the center exhibits central charge $c=8/5$, with scaling dimensions and energy spectra matching those of the orbifold Potts$^2$ theory.  

In particular, even at $J=0$ we numerically detect no effects of a putative $\epsilon^A \epsilon^B$ perturbation on critical exponents, as discussed in Sections \ref{subsec:central} and \ref{subsec:gSPT_SOP} above.
This suggests either that the cancellation mechanism discussed above indeed pins the critical line extremely close to $J=0$, or that the qutrit XX model realizes an unusually robust critical phase.  
It is interesting to note that Ref.~\onlinecite{Alavirad_2021} also found ``unnatural'' stability of a CFT (in that case with central charge $c=2$) to relevant perturbations in a similar class of models with antiferromagnetic $\mathbb{Z}_3\times \mathbb{Z}_3$-symmetric interactions. Determining whether $J \neq 0$ is strictly necessary to stabilize the critical line, and if so how large $J$ must be, is left to future high-precision numerics. Either way, the broader physical picture is unchanged.

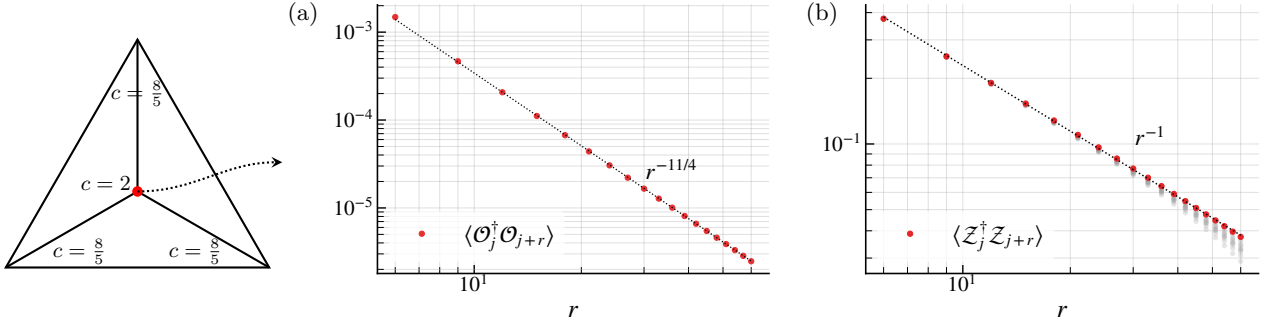
\begin{figure}
\begin{tikzpicture}
\node at (1.7,2){(a)};
\node at (8.6,2){(b)};
\node at (-0.5,0.2){
     \begin{tikzpicture}[scale=3.5, every node/.style={scale=0.9}]
            \path[use as bounding box] (-0.05,-0.05) rectangle (1.05,0.95);

            \coordinate (A) at (0,0);
            \coordinate (B) at (1,0);
            \coordinate (C) at (0.5,0.866);
            \coordinate (Center) at (0.5,0.289); 

            \draw[thick] (A) -- (B) -- (C) -- cycle;
            \draw[black, thick] (A) -- (Center);
            \draw[black, thick] (B) -- (Center);
            \draw[black, thick] (C) -- (Center);
            \fill[red] (Center) circle (0.02);

            \draw[overlay, densely dotted, line width=0.8pt,->,>=stealth]
                (Center) to[out=0,in=180] (1.05,0.40);

            \node at (0.38,0.32) {\small $c=2$};
            \node at (0.28,0.06) {\small $c=\tfrac{8}{5}$};
            \node at (0.72,0.06) {\small $c=\tfrac{8}{5}$};
            \node at (0.5,0.66) {\small $c=\tfrac{8}{5}$};
        \end{tikzpicture}
};
\node at (5,0){
\includegraphics[scale=0.4]{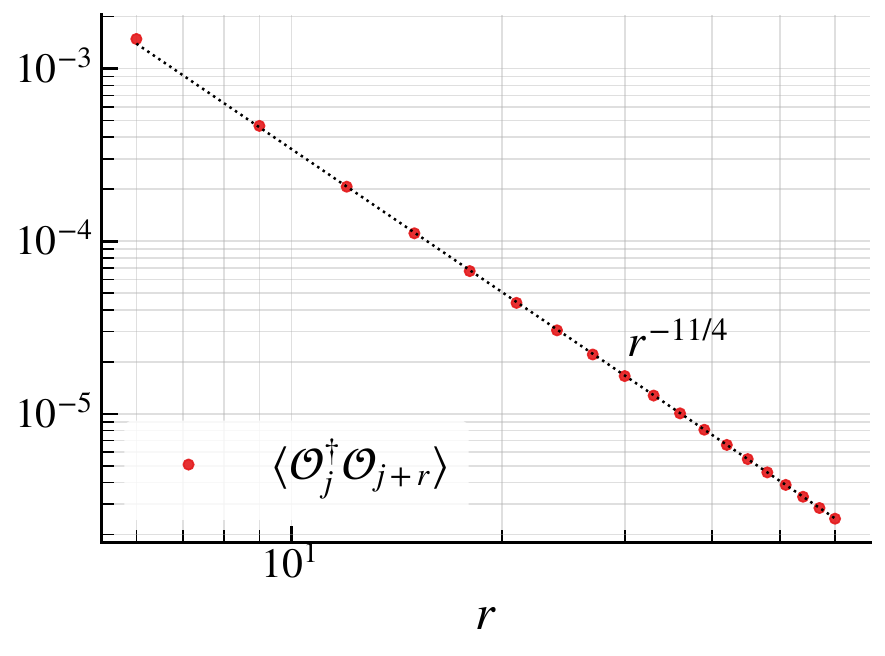}
};
\node at (11.5,0){
\includegraphics[scale=0.4]{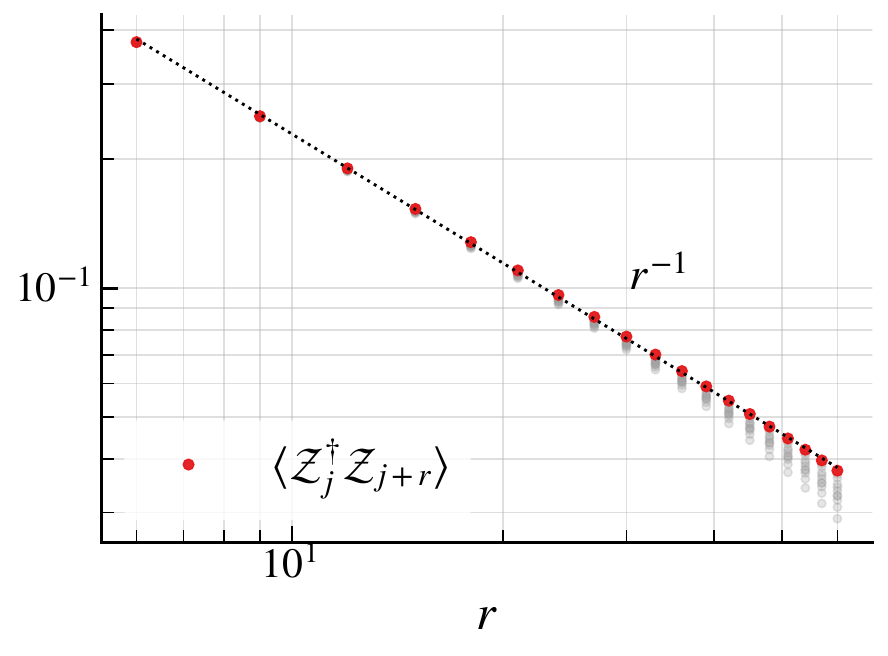}
};
\end{tikzpicture}
\caption{\textbf{Correlations at the $c=2$ multicritical point.}
At the $c=2$ multicritical theory at the center of the phase diagram (see Figure \ref{fig:triange_diagram}), we examine two-point functions $\langle \mathcal{O}_{j}^\dagger \mathcal{O}_{j+r} \rangle \propto r^{-2\Delta_{\mathcal{O}}}$ of local operators charged under single-site translation. (a) The operator $\mathcal O_j = X_{3j-1}^\dagger X_{3j} - X_{3j}^\dagger X_{3j+1} + (Z \leftrightarrow X) + h.c.$, which perturbs the multicritical theory into the SPT phase, has dimension $\Delta_{\mathcal{O}} \approx  11/8$. (b) The lightest local operator $\mathcal{Z}_j \equiv \sum_{a=0}^2 \omega^a Z_{3j+a}$ has dimension $\Delta_{\mathcal{Z}} \approx 1/2$. In Figure \ref{fig:string_charge_gapless_SPT} we also show the nonlocal twisted-sector operator scaling for this multicritical theory.
\label{fig:multicritical_decay}}
\end{figure}

\subsection{Multicriticality \label{subsec:multicrit}}

Most strikingly, our DMRG numerics suggests that the $c=8/5$ lines persist all the way to the center of the phase diagram. This shows that the anomalous theory there truly is a multicritical point between distinct gapless SPTs. In analogy with the discussion for the qubit model, the multicritical point with translation symmetry has an LSM anomaly forbidding a trivially gapped phase. 

Our entanglement-scaling numerics indicate that this multicritical point is a $c=2$ CFT, in agreement with earlier work \cite{Qin_2012, Alavirad_2021}. The precise location of this theory in the moduli space of $c=2$ CFTs remains an open question. Current evidence \cite{Alavirad_2021} suggests that it belongs to the continuous moduli space of marginal deformations of the $SU(3)_1$ Wess–Zumino–Witten theory. This identification is consistent with the  $\mathbb{Z}_3^3$ LSM anomaly, which forbids a non-degenerate short-range-entangled gapped ground state, and with conformal bootstrap bounds \cite{Lanzetta_2023} on-anomalous CFTs. In particular, we numerically find the lightest charged operator to have scaling dimension $1/2$, as shown in Fig.~\ref{fig:multicritical_decay}. This is consistent with the upper bound $2/3$ obtained via bootstrap for CFTs with this $\mathbb Z_3^3$ anomaly \cite{Lanzetta_2023}; while this bound is saturated for the ULS point at $J=1$, we see that for $J=0$ it is not.

In the underlying CFT, translation acts as an internal $\mathbb{Z}_3$ symmetry that permutes the three string order parameters from Eqn.~\ref{eq:string_order_params}, each a twist field with dimension $\Delta_{S} \approx 3/16$ as shown in Fig. \ref{fig:string_charge_gapless_SPT}. The two conjugate translation-breaking relevant perturbations into the SPT phases appear to have dimension $\Delta_a \approx 11/8$; this is consistent with the critical exponent of $\frac{3}{5} = \frac{2\Delta_{S}}{2-\Delta_{a}}$ of vanishing string order parameters in Figure \ref{fig:sptlro}. We can infer that for particular signs and linear combinations, these relevant perturbations drive RG flows into the three distinct $c=8/5$ orbifold Potts$^2$ theories
where gaplessness is preserved despite the anomaly being broken. 
The RG flow may bear some qualitative similarities with the well-known flow from the $c=1$  $U(1)_6$ CFT to $c=4/5$ three-state-Potts CFT ~\cite{Prakash_2025, Lecheminant_2002}, although we have not found a precise connection.
As an aside we remark that the lightest charged operator of the $c=8/5$ theory has dimension $\Delta^{\textrm{gSPT}}_{\text{min}} = 4/15 \approx 0.267$.
The theory lacks the $\mathbb{Z}_3^3$ anomaly, but it is interesting that its dimension is close to the bootstrap bound $\Delta^{\textrm{anomaly}}_{\text{min}}\leq 0.258$ for a putative anomalous $c=8/5$ theory~\cite{Lanzetta_2023,lanzetta}.
Testing our conjectured RG flow directly, for example via truncated conformal space methods~\cite{Yurov:1989yu, Hogervorst_2015}, would be an interesting direction for future work. As discussed in Sec.~\ref{subsec:stability}, concerns about the $\epsilon^A \epsilon^B$ perturbation at $J=0$ likely do not affect the overall picture, as there is a family of $c=2$ anomalous critical points at varying $J$~\cite{Alavirad_2021}.

\section{Local Symmetry Enrichment and Dipole Symmetry From Gauging}

In this last section, we discuss how dual (more precisely, gauged) formulations of the above model shed light on other types of symmetry-enriched quantum criticality.

\subsection{Local symmetry-enriched criticality}

A distinctive feature of certain critical systems is the existence of \emph{local symmetry-enriched criticalities} (SECs), in which global symmetries act in inequivalent ways on the \emph{local} operators of the CFT---even in the absence of spontaneous symmetry breaking. 
This enrichment has no analogue in nondegenerate gapped phases, for which symmetry only distinguishes the unique ground states in each twisted sector. 
At criticality, by contrast, the infrared theory retains detailed information about the charges of its low-energy excitations. 
If no (possibly emergent) symmetry relates different charge assignments, then they define genuinely distinct SECs.  Thus, local SECs represent new universality classes distinguished solely by how the same symmetry group acts on local operators. These distinctions can be regarded as different homomorphisms from the ultraviolet to infrared symmetry groups. 

For example, with $\mathbb{Z}_2 \times \mathbb{Z}_2^{\mathcal T}$ symmetry on a spin chain (generated by $\prod_j \sigma^x$ and complex conjugation in the $\sigma^z$ basis), the Ising spin field can be assigned to be $\mathbb{Z}_2^{\mathcal T}$-even or $\mathbb{Z}_2^{\mathcal T}$-odd. 
On the lattice, these correspond to
\begin{equation}
H_Z = -\sum_j \left(\sigma_j^z  \sigma_{j+1}^z + \sigma_j^x \right), 
\qquad
H_Y = -\sum_j \left( \sigma_j^y \sigma_{j+1}^y + \sigma_j^x \right),
\end{equation}
which realize inequivalent symmetry-enriched versions of the same Ising CFT \cite{Verresen_2021}.

For a case involving unitary symmetry and no gapped symmetry sectors, consider the $c=8/5$ Potts$^2$ CFT with protecting symmetry $G = \mathbb{Z}_3 \times \mathbb{Z}_3$, abstractly presented with independent generators $v$ and $w$ as  
\[
G = \mathbb{Z}_3 \times \mathbb{Z}_3 = \langle v,w \mid v^3=w^3=1,\; vw=wv \rangle .
\]
The four local operators with scaling dimension $\frac{2}{15}$ (namely $\sigma_A$, $\bar \sigma_A$, $\sigma_B$, $\bar \sigma_B$) naturally split into two doublets, each neutral under one proper subgroup of $G$ while carrying charge under another.  
This yields \emph{ inequivalent embeddings} of $G$ into two $\mathbb{Z}_3$ factors, and hence distinct local SECs, three of which we explore here. Explicitly, the lightest charged operators in these three cases are:
\begin{center}
\begin{tabular}{c|c}
Theory & Charges under $(v,w)$ \\ \hline
SEC-1 & $(\omega^{\pm1},1),\ (1,\omega^{\pm1})$ \\
SEC-2 & $(1,\omega^{\pm1}),\ (\omega^{\mp1},\omega^{\mp1})$ \\
SEC-3 & $(\omega^{\pm1},1),\ (\omega^{\mp1},\omega^{\mp1})$ 
\end{tabular}
\end{center}
No symmetry of the Potts$^2$ CFT, explicit or emergent, relates these three assignments.  
Hence they represent truly distinct universality classes.  
(This should be contrasted with the trivial case of the single Potts CFT with $\mathbb{Z}_3$ symmetry, where different charge assignments are rendered equivalent by emergent charge-conjugation symmetry in the IR.)

Although writing down such charge assignments is obvious, this allows us to then study the interplay between those distinct theories, especially in the form of multicritical transitions. 
The three SECs are related by automorphisms of $G$.  
These are generated by a `\emph{pivot}' operation (similar to Ref.~\onlinecite{pivot} for the topological case), which cyclically permutes the embeddings of $\mathbb{Z}_3$ subgroups into $G$.  
Concretely, the pivot acts as 
\begin{align}
U: &\ (v,w) \mapsto (w v^{-1}, v^{-1})
\end{align}
so that successive applications form a 3-cycle.
Unlike an SPT entangler, the pivot does \emph{not} commute with the protecting symmetry.  Together they generate a nonabelian group extension: the Heisenberg group mod $3$, $\mathrm{Heis}_3$.

At a multicritical point where all three SECs meet, the theory enjoys an emergent $\mathrm{Heis}_3$ symmetry.  
This symmetry is isomorphic to discrete $\mathbb{Z}_3$ dipole symmetry extended by translation.  
We now demonstrate that this multicritical structure of Potts$^2$ SECs is realized explicitly in a dipole-symmetric lattice model with translation acting as the pivot. In this construction, the three symmetry-enriched $c=8/5$ transitions meet at a $c=2$ theory with emergent $\mathrm{Heis}_3$ symmetry, exactly mirroring the multicriticality of twisted-sector-enriched gSPTs described earlier. The phenomenon follows from a general principle: gauging the subgroup of an anomalous symmetry at multicriticality naturally produces a nonabelian symmetry extension \cite{Bhardwaj_2018, Tachikawa_2020}. We show that dipole-symmetry-enriched multicriticality also arises when orbifold Potts$^2$ itself is viewed as a locally enriched transition between gapped dipole SPT phases.

\subsection{Model and phase diagram}

To make the discussion concrete, we now present a lattice model that realizes this multicritical structure. Consider the dipole-symmetric Hamiltonian
\begin{equation}
H_{\mathrm{dipole\ SSB}} = \sum_j a_{j \bmod 3} \left[ 
  X_j + Z_{j-1} Z_j Z_{j+1} 
  + J \bigl( Z_{j-1} Y_j Z_{j+1} + Z_{j-1} W_j Z_{j+1} \bigr) 
  + \text{h.c.} \right],
  \label{eq:dipoleSSB}
\end{equation}
which we will refer to as the \emph{dipole spontaneous symmetry breaking (SSB) model}.
This model has two independent $\mathbb{Z}_3$ symmetries: a dipole symmetry generated by the spatially-modulated operator $\prod_j X_j^j$, and a uniform global onsite symmetry $\prod_j X_j$, and we will see that this dipole symmetry is (partially) spontaneously broken \cite{pace2025spacetime, Pace_2025, Gorantla_2022, Cao_2024} in certain regions of our model.

\begin{figure}
\centering
\begin{tikzpicture}
\node at (-3.5,2.8) {(a)};
\node at (4.8,2.8) {(b)};
\node at (0,0){
\begin{tikzpicture}[scale=6]
\definecolor{bluePython}{rgb}{0.12, 0.47, 0.71}
\definecolor{tealPython}{rgb}{0.09, 0.74, 0.81}
\definecolor{greenPython}{rgb}{0.17, 0.63, 0.17}
\definecolor{orangePython}{rgb}{1.0, 0.5, 0.05}
\definecolor{gray1}{gray}{0.9}
\definecolor{gray2}{gray}{0.85}
\definecolor{gray3}{gray}{0.8}

\coordinate (A) at (0,0);
\coordinate (B) at (1,0);
\coordinate (C) at (0.5,0.866);
\coordinate (O) at (0.5,0.289);

\node[above] at (C) {\footnotesize $(a_0,a_1,a_2) = (1,0,0)$};
\node[below right] at (B) {\footnotesize$(0,1,0)$};
\node[below left] at (A) {\footnotesize$(0,0,1)$};

\fill[gray1] (A) -- (B) -- (O) -- cycle;
\fill[gray2] (B) -- (C) -- (O) -- cycle;
\fill[gray3] (C) -- (A) -- (O) -- cycle;

\draw[thick] (A) -- (B) -- (C) -- cycle;

\draw[thin] (A) -- (O);
\draw[thin] (B) -- (O);
\draw[thin] (C) -- (O);  

\node at (0.5,0.259-0.03) {\color{orangePython}$c=2$};

\node at (0.25,0.135) {$c=\frac{8}{5}$};
\node at (0.75,0.135) {$c=\frac{8}{5}$};
\node at (0.5,0.64)   {$c=\frac{8}{5}$};  

\node at (0.33,0.36) {SSB-1};
\node at (0.67,0.36) {SSB-2};
\node at (0.5,0.07)  {SSB-0};

\end{tikzpicture}
};
\node at (8,0){
\tdplotsetmaincoords{70}{110}
\begin{tikzpicture}
\node at (0,0){
    \begin{tikzpicture}[tdplot_main_coords, scale=4, transform shape,
                    line cap=round, line join=round, >=stealth]

    \coordinate (A) at (0,0,0);
    \coordinate (B) at ({1.3},0,0);
    \coordinate (C) at ({0.65},{1.1258},0);
    \coordinate (O) at ({0.65},{0.3757},0);

    \path (A) -- (O) coordinate[pos=0.2] (Ashift);
    \path (B) -- (O) coordinate[pos=0.2] (Bshift);
    \path (C) -- (O) coordinate[pos=0.2] (Cshift);

    \draw[line width=0.9pt] (A) -- (B) -- (C) -- cycle;
    \draw[line width=0.9pt] (A) -- (O);
    \draw[line width=0.9pt] (B) -- (O);
    \draw[line width=0.9pt] (C) -- (O);

    \foreach \p in {Ashift,Bshift,Cshift} {
      \draw[->, line width=1pt, red] (\p) -- ++(0,0,-0.20);
    }
    \draw[->, line width=1pt, orange] (O) -- ++(0,0,-0.20);

    \filldraw[tdplot_screen_coords, fill=gray!15, draw=none, opacity=0.8]
        (A) -- (B) -- (C) -- cycle;

    \foreach \p in {Ashift,Bshift,Cshift} {
      \draw[->, line width=1pt, blue]  (\p) -- ++(0,0,0.20);
    }
    \draw[->, line width=1pt, orange] (O) -- ++(0,0,0.20);

    \path (A) -- (B) coordinate[pos=0.5] (MAB);
    \path (B) -- (C) coordinate[pos=0.5] (MBC);
    \path (C) -- (A) coordinate[pos=0.5] (MCA);
    
    \path (O) -- (MAB) coordinate[pos=0.6666667] (SSB1); 
    \path (O) -- (MBC) coordinate[pos=0.6666667] (SSB0); 
    \path (O) -- (MCA) coordinate[pos=0.6666667] (SSB2); 
    
    \node[scale=0.15] at (SSB0) {SSB-0};
    \node[scale=0.15] at (SSB1) {SSB-1};
    \node[scale=0.15] at (SSB2) {SSB-2};

    \end{tikzpicture}
};
\draw[->] (-3,-2.2) -- (-3,2.2) node[above] {$\theta$};
\filldraw (-3,0) circle (2pt) node[left] {$\pi/4$};
\node[text=red] at (0,-2) {SPT$_1$};
\node[text=blue] at (0,2) {SPT$_\omega$};
\end{tikzpicture}

};
\end{tikzpicture}
\caption{\textbf{Gauging-related formulations: dipole symmetry-breaking and symmetry-enriched multiversality.} (a) The dipole SSB phase diagram (Eq.~\eqref{eq:dipoleSSB} with $J=0$) is obtained by gauging $\mathbb{Z}_3^X$ of the qutrit XX model. There are three different gapped phases spontaneously breaking the dipole symmetry, separated by inequivalent symmetry-enriched Potts$^2$ theories distinguished by the charges of their local operators. The symmetry enriched criticalities meet at a $c=2$ multicritical point with non-abelian symmetry. (b) Similarly, the dipole SPT model  \eqref{eq:dipoleSPT} arises from gauging $\mathbb{Z}_3^Y$. Here infinitessimally tuning the $\theta$ parameter leads to an orbifold Potts$^2$ transition from the dipole SPT phase to the trivial phase; in particular perturbing near the corners (red/blue arrows) immediately flows to two distinct gapped SPT phases. Between these same two gapped SPTs, there is a multiversality of three different symmetry-enriched orbifold Potts$^2$ CFTs, meeting at the $c=2$ multicritical point.}
\label{fig:dipole-phase}
\end{figure}
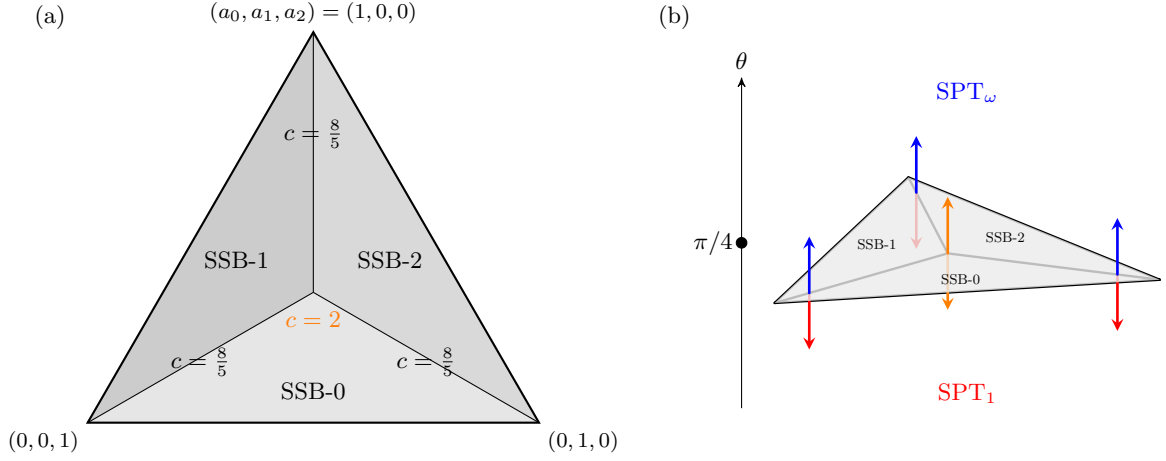

The dipole SSB model is precisely what one obtains by gauging the onsite $\mathbb Z^X_3$ symmetry of the qutrit XXYY chain.  
Under the Kramers–Wannier transformation
\[
X_j \mapsto Z_{j-1}^\dagger Z_{j}, \qquad Z_j \mapsto \prod_{k < j} X_k,
\]
the qutrit XXYY Hamiltonian is mapped to $H_{\mathrm{dipole\ SSB}}$.  
The phase diagram of $H_{\mathrm{dipole\ SSB}}$ is shown in Figure \ref{fig:dipole-phase}.  
It has the same triangular structure as the XX-chain diagram discussed earlier, with three gapped wedges (SSB phases) separated by critical segments ($c=8/5$) and a multicritical point ($c=2$) at the center.

The gapped SSB$-k$ phase breaks the global symmetry subgroup generated by $\prod X_j$ while preserving the global symmetry generated by $\prod_j X_j^{j-k}$ (for $k \in \{0,1,2\}$). Along the edges of the triangle (e.g., with $a_0=0$), operators $Z_j$ with $j\equiv k \pmod{3}$ commute with the Hamiltonian.  
They align as an order parameter that breaks the global symmetry $\prod_j X_j$, while preserving the modulated subgroup $\prod_j X_j^{j-k}$.  
The fixed-point wavefunction for the SSB-$k$ phase takes the form
\[
\bigotimes_j |w\rangle_{3j+k} \;\otimes\; 
\frac{1}{\sqrt{6}}\sum_{m,n} \mu_{m,n,w}\,|m,n\rangle_{3j+k+1,\,3j+k+2}, 
\]
where $w=0,1,2$ is chosen spontaneously (value of the order parameter) and $\mu_{m,n,w} \equiv m+n+w \pmod{3}$ takes values in $\{0,1,-1\}$.

Transitions between adjacent SSB phases are described by $c=8/5$ Potts$^2$ CFTs (without orbifold).  
At the corner limit of the SSB-$(k-1)$ to SSB-$(k+1)$ transition, the low-energy effective Hamiltonian is
\begin{equation}
H = -\frac{1}{2}\sum_{j \not\equiv k \bmod 3} \left( X_j + Z_j Z_{j+3}^\dagger + \text{h.c.} \right).
\end{equation}
The dipole symmetry acts differently on local spin operators $Z_j$ in each segment, so the three critical lines correspond to three inequivalent local SECs.

At the center of the diagram lies a $c=2$ CFT with emergent non-abelian $\mathrm{Heis}_3$ symmetry. 
This point unifies the three local SECs, reflecting the gauge-duality with gSPT multicriticality discussed in the rest of the paper.

\subsection{General remarks and duality: anomaly and symmetry extension}
\label{subsec:duality-general}

We now describe a general phenomenon: gauging an abelian subgroup of an anomalous symmetry with product group structure results in a dual symmetry that is a non-anomalous nontrivial group extension. This mechanism underlies the duality between LSM-type SPT physics and dipole symmetric models in our work.

We begin with an intuitive argument. Suppose we begin with a `type-III' anomalous symmetry group \cite{ElseNayak} of the form $U \times V \times W$, where $W$ is an abelian subgroup. The subgroups $U$ and $V$ commute on local operators, but in a $W$-twisted sector, $U$ acts nontrivially by permuting the $V$-charges attached to $W$-string operators. This defines a mixed anomaly.
After gauging $W$, these $W$-strings become local operators, and the action of the image $U'$ now permutes $V'$-charges of local operators. In this dual theory, the images $U'$ and $V'$ fail to commute on local operators:
\[
U'V'(U')^{-1}(V')^{-1} \in \widehat{W},
\]
where $\widehat{W}$ is the magnetic symmetry dual to $W$. This failure of commutativity indicates a non-abelian symmetry in the gauged theory. In particular, if $U$ is translation and gauging commutes with translation, then $U'$ still acts as translation, but now acts nontrivially on the dual symmetry $V'$, making $V'$ a \emph{spatially modulated symmetry}.

In our main example of the Qutrit XX model, we take 
$U=$ entangler (translation), $V=\mathbb{Z}_3^{Z}$, and $W=\mathbb{Z}_3^{X}$.  
Gauging $W$ maps $V$ to a dipole symmetry and results in a nontrivial commutator between $U$ and the dipole symmetry generators. 
The resulting symmetry group is the Heisenberg group mod~3, $\mathrm{Heis}_3$, explaining the modulated symmetry structure of the dipole SSB model.

This mechanism is formalized in Refs.~\onlinecite{Bhardwaj_2018, Tachikawa_2020} (with discussions on spatially-modulated symmetries in Refs.~\onlinecite{Aksoy_2024,Pace_2025, Pace_2025_2}). Suppose the anomalous symmetry is described by a direct product $Q \times W$, which fits into a short exact sequence
\begin{equation}
0 \to W \to Q \times W \to Q \to 0.
\end{equation}
Gauging $W$ results in a dual symmetry described by 
\begin{equation}
0 \to \widehat{W } \to \Gamma \to Q \equiv \Gamma/\widehat{W } \to 0.
\end{equation}
The anomaly cocycle of $Q \times W$ precisely corresponds to the nontrivial symmetry extension $\Gamma$ defining the multicritical point, with the explicit relation given in Ref.~\onlinecite{Tachikawa_2020}.

In our study of phase transitions, $Q$ further factorizes into $Q = U\times V$, such that $V \times W$ is \textit{non-anomalous}. Then $U$ plays the role of the pivot.  Once the pivot $U$ is explicitly broken away from the multicritical point, there is no anomalous symmetry in the pre-gauged theory and no non-abelian symmetry in the gauged theory.  What remains in the pre-gauged theory 
is a set of inequivalent SPT phases protected by $V \times W$. These SPTs are related by the action of the broken $U$, and are distinguished by their string order parameters.  
Upon gauging $W$, each string order parameter becomes a local order parameter, so that the $V \times W$ SPT phases map directly to dipole SSB phases. Thus, away from the multicritical point the duality is: SPT $\leftrightarrow$ SSB, string order $\leftrightarrow$ local order, and gSPT transitions $\leftrightarrow$ local SEC transitions.

\subsection{Symmetry-enriched multiversality of dipole SPT transitions}
\label{subsec:dipole-spt}

In this last section, we will discuss how symmetries and automorphisms also provide new insight into \emph{multiversality} \cite{Bi19,Prakash23}, which refers to the fact that the critical theory between two gapped phases is \emph{not} uniquely determined by the nature of those phases.  
In particular, we point out that while gapped SPTs are invariant under cohomology-preserving automorphisms, their critical points need not be.
Thus, the same pair of SPT phases can be connected by multiple inequivalent CFTs, a phenomenon we call \emph{symmetry-enriched multiversality}.  
This naturally motivates the study of their interplay and possible multicriticality.

For example, $\mathbb{Z}_3\times\mathbb{Z}_3$ SPT phases are known to be connected by an orbifold Potts$^2$ CFT, as we explored in Section \ref{sec:gapless}. However, the same two SPTs can be connected by multiple inequivalent local enrichments of this CFT. We can explicitly realize a $c=2$ multicriticality of these SECs in a dipole-symmetric setup with the following Hamiltonian:
\begin{equation}
H_{\textrm{Dipole SPT}}(\theta) = \sum_j a_{j\bmod 3} \Bigl( \cos(\theta)\, X_j + \sin(\theta)\, Z_{j-1} Y_j Z_{j+1}  + \textrm{h.c.}\Bigr),
\label{eq:dipoleSPT}
\end{equation}
where as before we focus on $a_j \geq 0$. Each individual SEC can be regarded as a transition between dipolar SPT phases~\cite{han2023topologicalquantumchainsprotected, lam2023classificationdipolarsymmetryprotectedtopological, pace2025spacetime,  kim2025noninvertiblesymmetrytopologicalholography, bulmash2025defectnetworkstopologicalphases}. 

At $\theta=\pi/4$, this Hamiltonian is dual to the qutrit XX model upon gauging the diagonal $\mathbb{Z}_3^Y$ subgroup generated by $\prod_j Y_j$; the resulting phase diagram contains the same dipole SSB phases as before but separated by \textit{orbifold} Potts$^2$ transitions. Near the corners, tuning $\theta$ downward or upward from $\pi/4$ perturbs the CFT into the trivial phase or dipole SPT phase~\cite{han2023topologicalquantumchainsprotected}.
Thus between the same two SPT phases, we realize three distinct symmetry-enriched transitions for which translation acts like the group automorphism that pivots between them. 
As a remark, replacing $X_j$ with $Z_{j-1} W_j Z_{j+1} $ in the Hamiltonian would instead represent a transition from one nontrivial dipole SPT to the other. 

The three orbifold Potts$^2$ lines meet at a $c=2$ multicritical point with $\mathrm{Heis}_3$ symmetry.  Unlike at the modulated corner, here translation symmetry stabilizes an extended $c=2$ region: tuning $\theta$ away from $\pi/4$ does not immediately gap out the system \cite{Qin_2012}.  Numerically, this $c=2$ phase persists approximately for $-0.1\pi \lesssim \theta \lesssim 0.6\pi$.  For $\theta\gtrsim0.6\pi$ the system enters the trivial phase, while for $\theta\lesssim-0.1\pi$ it realizes the dipole SPT phase of Ref.~\onlinecite{han2023topologicalquantumchainsprotected}.

In the limit $\theta \ll 1$, the low-energy Hilbert space reduces to effective qubits where the second term acts as a three-site spin raising/lowering operator.
Remarkably, this coincides with a special limit of the Bose–Hubbard model with modulated symmetry studied by Ref.~\onlinecite{Sala_2024}.  
Their Hamiltonian,
\begin{equation}
H_{\textrm{Bose Hubbard}} = -J \sum_j b_{j-1} b^\dagger_j b_{j+1}+ {\rm h.c.} + \frac{U}{2}\sum_j (n_j - \mu/U)^2
\end{equation}
reduces exactly to the same effective qubit model with appropriate relabelings when $\mu/U=1/2$ and $U/J\to\infty$. 

This connection suggests a broader correspondence between local symmetry-enriched criticality transitions and modulated-symmetry models, and motivates future study of their full phase diagrams.

\section{Conclusion and Outlook}

Our work establishes an example of multicriticality between inequivalent \emph{purely gapless} SPT phases with a unitary protecting symmetry.
We used the LSM anomaly as a guideline to proposing the model, namely the (trimerized) qutrit XX chain. We demonstrated that three distinct $c=\tfrac{8}{5}$ orbifold Potts$^2$ gapless SPTs collide at a multicritical point described by a $c=2$ conformal field theory. Translation acts as an entangler, cyclically permuting the gapless SPTs, and at the multicritical point it gives rise to a mixed anomaly with the global $\mathbb{Z}_3\times\mathbb{Z}_3$ symmetry. We verified this realization of multicriticality through analytic mappings and numerical simulations.
We further showed how gauging maps this multicriticality into one of dipole-symmetric models, where it corresponds to a $c=2$ theory with non-abelian $\mathrm{Heis}_3$ symmetry unifying distinct locally symmetry-enriched criticalities. 

Looking forward, several directions remain open.
It would be interesting to determine the precise identity of the multicritical theory within the moduli space of $c=2$ CFTs, and to understand the RG flows emanating from it, possibly with the help of truncated conformal space approach \cite{Yurov:1989yu, Hogervorst_2015}. Moreover, our work raises a peculiar puzzle of understanding the surprisingly strong stability of the $c=8/5$ theory observed numerically throughout the SPT transition in the qutrit XX model---it remains to be seen whether a small nonzero value of $J$ is necessary to stay along the $c=\frac{8}{5}$ line, or whether there is an unusual stability similar to the case of $c=2$ in Ref.~\onlinecite{Alavirad_2021}. Varying $J$ and breaking Hadamard symmetry could reveal additional phases and phase transitions.
More broadly, it would be valuable to systematically classify multicritical points connecting distinct gapless SPTs, including intrinsically gapless cases and higher-dimensional generalizations.
Our work connects to recent discussions of fundamental constraints on CFTs, with several promising directions especially relating to anomalies and SPT transitions ~\cite{Verresen17, Tsui_2017,
cordova2022symmetryenrichedctheorems}. For instance, Ref.~\onlinecite{Alavirad_2021} proposed that a $\mathbb{Z}_d^3$-anomalous CFT should satisfy central charge $c \geq d-1$. Meanwhile, the ``$c$--$d$ conjecture''~\cite{Latorre_2024} suggests that a nearest-neighbor translationally-invariant qudit critical chain is bounded by $c \le d-1$. Our model saturates both conjectural bounds.

Although this paper focuses on bulk properties, gapless SPTs are also known to host rich boundary physics. An important question is the fate of the stable edge modes~\cite{Prembabu_2022} of gSPT$_{\omega, \overline{\omega}}$ as the theory is tuned towards multicriticality. Our phase diagram further motivates asking whether the boundary deconfined quantum critical point identified in Ref.~\onlinecite{Prembabu_2022} can be interpreted as an RG interface of the multicritical theory with a gapless SPT on one side and a trivial bulk on the other \cite{verresen2020topologyedgestatessurvive}. Other RG interface setups (using our rich phase diagram) may help understand defects between distinct gapless SPTs, which are expected to exhibit nontrivial constraints \cite{prembabu}.

Finally, our \emph{nearest-neighbor} qutrit structure suggests that experimental realizations in cold-atom quantum simulators or related platforms may be within reach \cite{QSim}. While this could already be of interest for the gapped SPT phases in our model, it would be especially interesting for the gapless SPT phases which arise as continuous phase transitions between multiple nontrivial gapped SPT phases. In particular, our gSPT$_{\omega,\bar \omega}$ model is a nearest-neighbor chain which is critical in the bulk but has robust edge modes, despite a complete absence of gapped degrees of freedom. This makes it an ideal platform for studying the interplay between topology and criticality. 

\section*{Acknowledgments} 
The authors thank Omer Aksoy, Takamasa Ando, Maissam Barkeshli, Nick Jones, Ho Tat Lam, Ryan Lanzetta, Olexei Motrunich, Masaki Oshikawa, Sal Pace, Abhishodh Prakash, Pablo Sala, Madhav Sinha, Ryan Thorngren, Saran Vijayan and Ashvin Vishwanath for useful discussions. R.V. also thanks Julian Bibo and Frank Pollmann for collaboration on a related work. S.P. thanks K. Nirmala for support. S.P. was supported by the National Science Foundation grant NSF-DMR 2220703. For the initial stages of this work, R.V. was supported by the Simons Collaboration on Ultra-Quantum Matter, which is a grant from the Simons Foundation (651440, Ashvin Vishwanath). This research was supported in part by grant NSF PHY-2309135 to the Kavli Institute for Theoretical Physics (KITP) while attending the `Generalized Symmetries' program.

\bibliographystyle{unsrt} 
\bibliography{references} 

\appendix

\section{Numerical Details}
\label{app:numerics}
We use infinite DMRG (iDMRG) with a unit cell of three sites, implemented using the \textsc{TeNPy} library~\cite{Hauschild_2018}. The ground state is represented as an infinite matrix product state with period-3 tensors $A_1$, $A_2$, and $A_3$. The state takes the form
\[
|\psi\rangle = \sum_{\{s_j\}} \mathrm{Tr}\left[\cdots A_1^{s_1} A_2^{s_2} A_3^{s_3} A_1^{s_4} \cdots \right]\,|\{s_j\}\rangle,
\]
where $A_k^s$ denotes the $\chi \times \chi$ matrix slice of $A_k$ at physical index $s \in \{1, \dots, d\}$.

The corresponding tensor network diagram is:

\[
\begin{tikzpicture}[baseline={([yshift=-.5ex]current bounding box.center)}, scale=1]

\node at (-1.0,0.25) {$\cdots$};
\draw[-] (-0.7,0.25) -- (0,0.25);

\foreach \i/\lbl in {0/A_1, 1/A_2, 2/A_3, 3/A_1, 4/A_2, 5/A_3, 6/A_1, 7/A_2, 8/A_3} {
  \draw[fill=gray!20, rounded corners=2pt] (\i,0) rectangle +(0.8,0.5);
  \node at (\i+0.4,0.25) {$\lbl$};
  \draw[-] (\i+0.4,0) -- (\i+0.4,-0.5);
}

\foreach \i in {0,...,7} {
  \draw[-] (\i+0.8,0.25) -- (\i+1,0.25);
}

\draw[-] (8.8,0.25) -- (9.5,0.25);
\node at (9.8,0.25) {$\cdots$};

\end{tikzpicture}
\]

\vspace{1em}

We contract tensors to compute expectation values. For example, $\langle \mathcal{O}_{3j} \mathcal{O'}_{3j+2}\rangle$ can be graphically evaluated as:

\[
\begin{tikzpicture}[baseline={([yshift=-.5ex]current bounding box.center)}, scale=1]

\foreach \i/\lbl in {0/A_1, 1/A_2, 2/A_3, 3/A_1, 4/A_2, 5/A_3, 6/A_1, 7/A_2, 8/A_3} {
  \draw[fill=gray!20, rounded corners=2pt] (\i,1.0) rectangle +(0.8,0.5);
  \node at (\i+0.4,1.25) {$\lbl$};
}

\foreach \i/\lbl in {0/A_1^*, 1/A_2^*, 2/A_3^*, 3/A_1^*, 4/A_2^*, 5/A_3^*, 6/A_1^*, 7/A_2^*, 8/A_3^*} {
  \draw[fill=gray!20, rounded corners=2pt] (\i,0) rectangle +(0.8,0.5);
  \node at (\i+0.4,0.25) {$\lbl$};
}

\foreach \i in {-1,0,1,2,3,4,5,6,7,8} {
  \draw[-] (\i+0.8,1.25) -- (\i+1,1.25); 
  \draw[-] (\i+0.8,0.25) -- (\i+1,0.25); 
}

\foreach \i in {0,...,8} {
  \draw[-] (\i+0.4,1.0) -- (\i+0.4,0.5);
}

\foreach \i in {2} {
  \draw[fill=white] (\i+0.25,0.65) rectangle +(0.3,0.2);
  \node at (\i+0.4,0.75) {\tiny $\mathcal{O}$};
}

\foreach \i in {4} {
  \draw[fill=white] (\i+0.25,0.65) rectangle +(0.3,0.2);
  \node at (\i+0.4,0.75) {\tiny $\mathcal{O'}$};
}

\node at (-0.5,1.25) {$\cdots$};
\node at (-0.5,0.25) {$\cdots$};
\node at (9.5,1.25) {$\cdots$};
\node at (9.5,0.25) {$\cdots$};

\end{tikzpicture}
\]

\subsection{Extraction of gapped long range order}

We are interested in the long range correlations such as $\langle \prod_{j=3m+1}^{3n} X_j\rangle$, shown graphically:

\[
\begin{tikzpicture}[baseline={([yshift=-.5ex]current bounding box.center)}, scale=1]

\node at (-0.5,1.25) {$\cdots$};
\node at (-0.5,0.25) {$\cdots$};

\draw[-] (-0.2,1.25) -- (0,1.25); 
\draw[-] (-0.2,0.25) -- (0,0.25); 

\foreach \i/\lbl in {0/A_3, 1/A_1, 2/A_2, 3/A_3, 4/A_1} {
  \draw[fill=gray!20, rounded corners=2pt] (\i,1.0) rectangle +(0.8,0.5);
  \node at (\i+0.4,1.25) {$\lbl$};
}
\foreach \i/\lbl in {0/A_3^*, 1/A_1^*, 2/A_2^*, 3/A_3^*, 4/A_1^*} {
  \draw[fill=gray!20, rounded corners=2pt] (\i,0) rectangle +(0.8,0.5);
  \node at (\i+0.4,0.25) {$\lbl$};
}


\foreach \i in {0} {
  \draw[-] (\i+0.4,1.0) -- (\i+0.4,0.5);
}

\foreach \i in {1,2,3,4} {
  \draw[-] (\i+0.4,1.0) -- (\i+0.4,0.9);
  \draw[fill=white] (\i+0.25,0.65) rectangle +(0.3,0.2);
  \node at (\i+0.4,0.75) {\tiny $X$};
  \draw[-] (\i+0.4,0.65) -- (\i+0.4,0.5);
}

\foreach \i in {0,1,2,3} {
  \draw[-] (\i+0.8,1.25) -- (\i+1,1.25); 
  \draw[-] (\i+0.8,0.25) -- (\i+1,0.25); 
}


\draw[-] (4.8,1.25) -- (5.0,1.25);
\draw[-] (4.8,0.25) -- (5.0,0.25);

\node at (5.5,1.25) {$\cdots$};
\node at (5.5,0.25) {$\cdots$};
\node at (5.5,0.75) {$\cdots$};

\draw[-] (5.8,1.25) -- (6.0,1.25);
\draw[-] (5.8,0.25) -- (6.0,0.25);

\foreach \i/\lbl in {6/A_2, 7/A_3, 8/A_1, 9/A_2} {
  \draw[fill=gray!20, rounded corners=2pt] (\i,1.0) rectangle +(0.8,0.5);
  \node at (\i+0.4,1.25) {$\lbl$};
}
\foreach \i/\lbl in {6/A_2^*, 7/A_3^*, 8/A_1^*, 9/A_2^*} {
  \draw[fill=gray!20, rounded corners=2pt] (\i,0) rectangle +(0.8,0.5);
  \node at (\i+0.4,0.25) {$\lbl$};
}
\foreach \i in {6,7} {
  \draw[-] (\i+0.4,1.0) -- (\i+0.4,0.9);
  \draw[fill=white] (\i+0.25,0.65) rectangle +(0.3,0.2);
  \node at (\i+0.4,0.75) {\tiny $X$};
  \draw[-] (\i+0.4,0.65) -- (\i+0.4,0.5);
}

\foreach \i in {8,9} {
  \draw[-] (\i+0.4,1.0) -- (\i+0.4,0.5);
}

\draw[-] (6.8,1.25) -- (7,1.25);
\draw[-] (6.8,0.25) -- (7,0.25);
\draw[-] (7.8,1.25) -- (8,1.25);
\draw[-] (7.8,0.25) -- (8,0.25);
\draw[-] (8.8,1.25) -- (9,1.25);
\draw[-] (8.8,0.25) -- (9,0.25);

\draw[-] (9.8,1.25) -- (10.0,1.25);
\draw[-] (9.8,0.25) -- (10.0,0.25);
\node at (10.3,1.25) {$\cdots$};
\node at (10.3,0.25) {$\cdots$};

\end{tikzpicture}
\]

The transfer matrix formalism makes it possible to compute the infinite-length
long-range order (LRO) of string operators at finite bond dimension $\chi$~\cite{Pollmann_2012_detection}.
Concretely, we can introduces the unit cell transfer matrices
\begin{equation}
T^{(0)} \;=\;
\begin{tikzpicture}[baseline={([yshift=-.5ex]current bounding box.center)}, scale=0.9]
  \foreach \i/\lbl in {1/A_1, 2/A_2, 3/A_3} {
    \draw[fill=gray!20, rounded corners=2pt] (\i,1.0) rectangle +(0.8,0.5);
    \node at (\i+0.4,1.25) {$\lbl$};
  }
  \foreach \i/\lbl in {1/A_1^*, 2/A_2^*, 3/A_3^*} {
    \draw[fill=gray!20, rounded corners=2pt] (\i,0) rectangle +(0.8,0.5);
    \node at (\i+0.4,0.25) {$\lbl$};
  }
  \foreach \i in {1,2,3} {
    \draw[-] (\i+0.4,1.0) -- (\i+0.4,0.5);
  }
  \foreach \i in {0,1,2,3} {
    \draw[-] (\i+0.8,1.25) -- (\i+1,1.25);
    \draw[-] (\i+0.8,0.25) -- (\i+1,0.25);
  }
\end{tikzpicture}
\qquad , \qquad
T^{(0)}_{X} \;=\;
\begin{tikzpicture}[baseline={([yshift=-.5ex]current bounding box.center)}, scale=0.9]
  \foreach \i/\lbl in {1/A_1, 2/A_2, 3/A_3} {
    \draw[fill=gray!20, rounded corners=2pt] (\i,1.0) rectangle +(0.8,0.5);
    \node at (\i+0.4,1.25) {$\lbl$};
  }
  \foreach \i/\lbl in {1/A_1^*, 2/A_2^*, 3/A_3^*} {
    \draw[fill=gray!20, rounded corners=2pt] (\i,0) rectangle +(0.8,0.5);
    \node at (\i+0.4,0.25) {$\lbl$};
  }
  \foreach \i in {1,2,3} {
    \draw[-] (\i+0.4,1.0) -- (\i+0.4,0.9);
    \draw[fill=white] (\i+0.25,0.65) rectangle +(0.3,0.2);
    \node at (\i+0.4,0.75) {\tiny $X$};
    \draw[-] (\i+0.4,0.65) -- (\i+0.4,0.5);
  }
  \foreach \i in {0,1,2,3} {
    \draw[-] (\i+0.8,1.25) -- (\i+1,1.25);
    \draw[-] (\i+0.8,0.25) -- (\i+1,0.25);
  }
\end{tikzpicture}
\end{equation}
acting on the doubled virtual space of dimension $\chi^2$.

Each transfer matrix has a single dominant eigenvalue of modulus one (since we are targeting non-symmetry-breaking phases of matter),
while all subleading eigenvalues have smaller magnitude.  
This property guarantees that in the infinite-length limit, products of transfer
matrices simplify dramatically: the action of $T^{(0)}$ or $T^{(0)}_X$ becomes
equivalent to a projection onto the corresponding leading eigenspace.

We denote the dominant left and right eigenvectors of $T^{(0)}$ by
$\langle L^{(0)}|$ and $|R^{(0)}\rangle$, and those of $T^{(0)}_X$ by
$\langle L^{(0)}_X|$ and $|R^{(0)}_X\rangle$.  
All these vectors live in the $\chi^2$-dimensional doubled virtual space,
and we normalize them such that
\[
\langle L^{(0)} | R^{(0)} \rangle = 1, 
\qquad
\langle L^{(0)}_X | R^{(0)}_X \rangle = 1.
\]
(Note that in the ``right canonical form'' convention, $\ket{R}$ is proportional to the $\chi \times \chi$ identity matrix.) With this setup, correlation functions and string order parameters at finite bond dimension $\chi$ can be extracted directly from overlaps involving these
leading eigenvectors:

\[
\lim_{|m-n|\to\infty}\langle \prod_{j=3m+1}^{3n}X_i\rangle = \frac{\langle L^{(0)} | R^{(0)} _X\rangle \langle L^{(0)} _X | R^{(0)} \rangle}  
{\langle L^{(0)} | R^{(0)}  \rangle \langle L^{(0)} _X | R^{(0)} _X\rangle}  
\]

For the other string order parameters we can use the transfer matrices with the tensors cyclically permuted; these have exactly the same eigenvalues but different eigenvectors. 

\begin{equation}
T^{(1)} \;=\;
\begin{tikzpicture}[baseline={([yshift=-.5ex]current bounding box.center)}, scale=0.9]
  \foreach \i/\lbl in {1/A_2, 2/A_3, 3/A_1} {
    \draw[fill=gray!20, rounded corners=2pt] (\i,1.0) rectangle +(0.8,0.5);
    \node at (\i+0.4,1.25) {$\lbl$};
  }
  \foreach \i/\lbl in {1/A_2^*, 2/A_3^*, 3/A_1^*} {
    \draw[fill=gray!20, rounded corners=2pt] (\i,0) rectangle +(0.8,0.5);
    \node at (\i+0.4,0.25) {$\lbl$};
  }
  \foreach \i in {1,2,3} {
    \draw[-] (\i+0.4,1.0) -- (\i+0.4,0.5);
  }
  \foreach \i in {0,1,2,3} {
    \draw[-] (\i+0.8,1.25) -- (\i+1,1.25);
    \draw[-] (\i+0.8,0.25) -- (\i+1,0.25);
  }
\end{tikzpicture}
\qquad , \qquad
T^{(1)}_{X} \;=\;
\begin{tikzpicture}[baseline={([yshift=-.5ex]current bounding box.center)}, scale=0.9]
  \foreach \i/\lbl in {1/A_2, 2/A_3, 3/A_1} {
    \draw[fill=gray!20, rounded corners=2pt] (\i,1.0) rectangle +(0.8,0.5);
    \node at (\i+0.4,1.25) {$\lbl$};
  }
  \foreach \i/\lbl in {1/A_2^*, 2/A_3^*, 3/A_1^*} {
    \draw[fill=gray!20, rounded corners=2pt] (\i,0) rectangle +(0.8,0.5);
    \node at (\i+0.4,0.25) {$\lbl$};
  }
  \foreach \i in {1,2,3} {
    \draw[-] (\i+0.4,1.0) -- (\i+0.4,0.9);
    \draw[fill=white] (\i+0.25,0.65) rectangle +(0.3,0.2);
    \node at (\i+0.4,0.75) {\tiny $X$};
    \draw[-] (\i+0.4,0.65) -- (\i+0.4,0.5);
  }
  \foreach \i in {0,1,2,3} {
    \draw[-] (\i+0.8,1.25) -- (\i+1,1.25);
    \draw[-] (\i+0.8,0.25) -- (\i+1,0.25);
  }
\end{tikzpicture}
\end{equation}

\subsection{Convergence in $\chi$}

In matrix product simulations, the accuracy of ground state is controlled by the bond dimension $\chi$.  For gapped phases far from criticality, relatively small values of $\chi$ are already sufficient to capture the exact ground state.  For instance, at $H = H_0 + 0.325 H_1 + 0.675 H_2$, we find that $\chi=535$ fully saturates the ground-state properties. 

\begin{figure}[t]
    \centering
    \begin{tikzpicture}
        \node at (-4.0,3.1) {{(a)}};
        \node at (5.1,3.1) {{(b)}};

        \node at (0,0) {
            \includegraphics[width=0.49\linewidth]{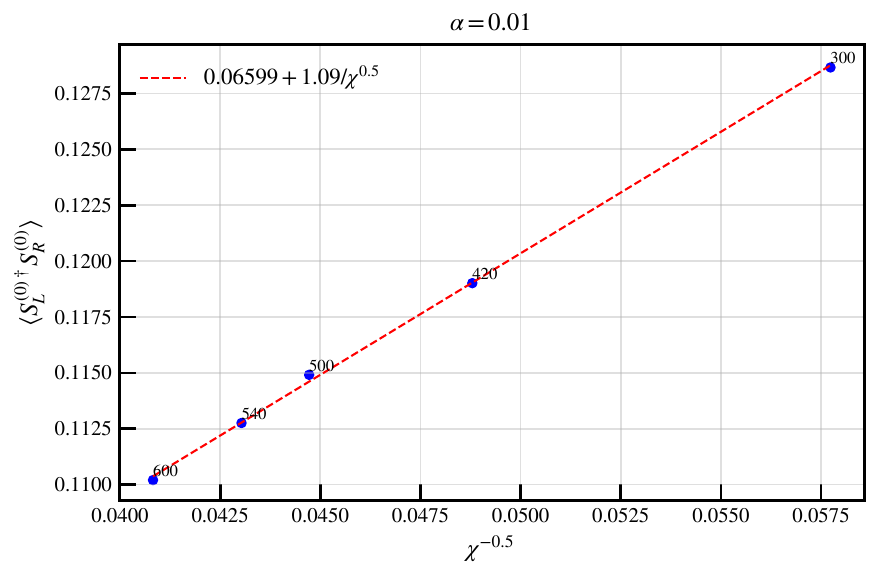}
        };

        \node at (9.0,0) {
            \includegraphics[width=0.49\linewidth]{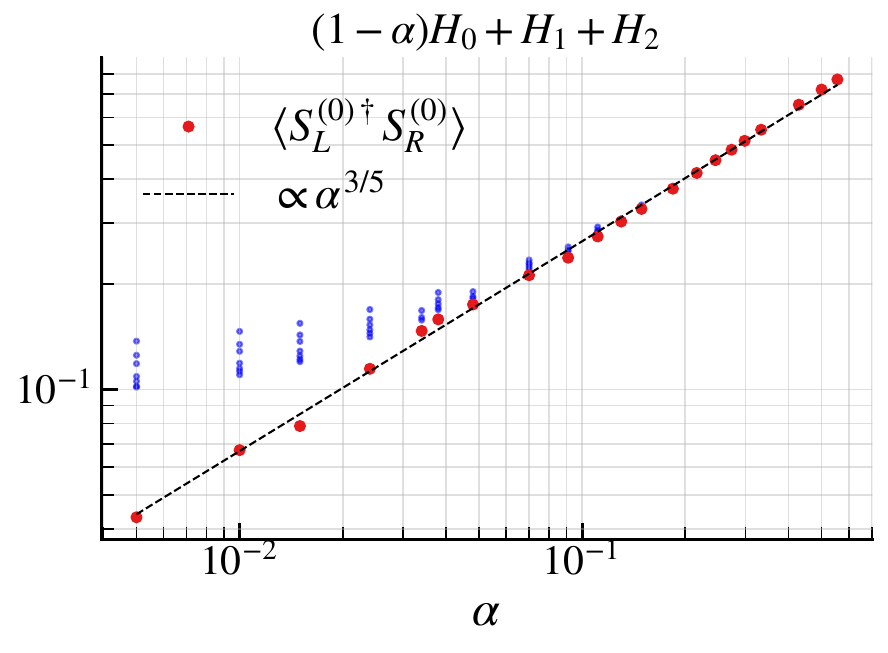}
        };
    \end{tikzpicture}

    \caption{
    \textbf{Extrapolating observables in the $\boldsymbol{\chi \to \infty}$ limit.}
    (a) We extract the long-range order (LRO) from finite-$\chi$ data by fitting to the form
    $f_\chi = f_\infty + \mathrm{const.}/\chi^p$, with the exponent $p$ chosen to optimize fit.
    This example shows the extrapolation for the string order parameter in SPT$_1$ near the multicritical point.
    (b) The extrapolated value $f_\infty = 0.06599$ reasonably matches the scaling predictions from CFT.
    }
    \label{fig:chi_convergence}
\end{figure}

The situation is different at or near critical points, the focus of our work. Gapless ground states have infinite correlation length and logarithmically-growing entanglement, so they cannot be exactly represented at finite~$\chi$. Finite $\chi$ provides an approximate variational state with a fictitious correlation length $\propto \chi^\kappa$ where $\kappa ={\frac{6}{c\left(\sqrt{\frac{12}{c}}+1\right)}}$ for a CFT with central charge $c$~\cite{Pollmann_2009}. Even in gapped phases close to the critical point, the diverging correlation length is typically too large to be practically reproduced by finite bond dimension.  In these cases one must instead extrapolate finite-$\chi$ data to the $\chi \to \infty$ limit.  Throughout this work we adopted a simple scheme for such extrapolations.  
For a physical observable, estimates from finite~$\chi$ simulations $f_\chi$ were fit to the form \begin{equation}
f_\chi = f_\infty + \frac{\mathrm{const.}}{\chi^p} \, ,
\end{equation}
where the exponent $p$ is approximately chosen to optimize a linear fit. Higher weight was given to higher-$\chi$ points in the linear regression. Converged data was generally not highly sensitive to the precise value of $p$. For instance we show this method in Figure~\ref{fig:chi_convergence} for computing the long range order in a string order parameter slightly away from the multicritical point. Although the true physical long range order is nowhere near the estimates computed at $\chi\leq 600$, we can reliably extract it by extrapolation.

\section{Details of Schrieffer-Wolff Projection}
\label{app:SW}

\subsection{Potts$^2$ language}

For the analytic calculations it will be most convenient to use the language of the Potts$^2$ model, rather than the orbifold Potts$^2$ model. 
The variables on non-auxiliary sites can be conveniently described in terms of Potts$^2$ variables, upon gauging a $\mathbb{Z}_3$ subgroup following a precise procedure in Ref.~\onlinecite{Prembabu_2022}.
The result after gauging (and site re-labeling) is 

\begin{equation}
\begin{split}
H_0^{\textrm{Potts}} &= \sum_j  \color{red} Z_{j+\frac{1}{2}} \color{black}  +  \color{red} X_{j+\frac{1}{2}} \color{black}  + h.c.\\
H_1^{\textrm{Potts}} &= \sum_j  \color{red} Z_{j+\frac{1}{2}} \color{black}  X^{B\dagger}_{j+1} +  Z_{j}^{A}\color{red} X_{j+\frac{1}{2}} \color{black} Z^{A\dagger}_{j+1} + h.c.\\
H_2^{\textrm{Potts}}&= \sum_j X^{A}_{j} \color{red} Z_{j+\frac{1}{2}} \color{black}  + Z^{B}_{j}\color{red} X_{j+\frac{1}{2}} \color{black} Z^{B\dagger}_{j+1}  + h.c.    
\end{split}
\end{equation}

Hereafter in this appendix we will drop the label ``Potts'' for convenience. Furthermore, for convenience we can write $H_1+H_2$ as 

\[H_1+H_2 = V = \sum_{j} V_{j+\frac{1}{2}} \equiv \sum_{j} \sum_{\alpha = 0}^3 \sigma_{j+\frac{1}{2}}^\alpha V^\alpha_{j+\frac{1}{2}}\]

Where $\sigma^{0},\sigma^{1}, \sigma^{2}, \sigma^{3}$ refer to the clock matrices $X, Z, X^\dagger, Z^\dagger$ respectively, and $V^\alpha_{j+\frac{1}{2}}$ only has support on the integer sites adjacent to $j+\frac{1}{2}$.
That is, $V^{(0)}_{j+\frac{1}{2}} =  Z^A_{j} Z^{A\dagger}_{j+1} + Z^B_{j} Z^{B\dagger}_{j+1}$,  $V^{(1)}_{j+\frac{1}{2}} =  X^{A}_{j}+X^{B\dagger}_{j+1} $, and $V^{\alpha+2}_{j+\frac{1}{2}} = V^{\alpha \dagger}_{j+\frac{1}{2}}$

The Hadamard $\mathbb{Z}_4$ symmetry of the original qutrit XXYY model shifts $\alpha \to \alpha+1$. Furthermore, we introduce shorthand notation for local lattice operators of various Hadamard charges, which in turn may be expressed as a sum of operators from the $A$ and $B$ chains each with a known correspondence to CFT fields \cite{Mong_2014} and leading scaling dimensions $\Delta$. Here, $\Phi_{X\overline{\epsilon}}$ and $\Phi_{\epsilon \overline{X}}$  refer to the charge-conjugaton-odd $W$-algebra descendants of dimension $9/5$ of the thermal field $\epsilon$, while $T(z)$ is the holomorphic stress-energy tensor.

\begin{center}
\begin{tabular}{c|c|c|l}
\toprule
Shorthand & Lattice definition & CFT correspondence & $\Delta$ \\
\hline
$T$ & $\sum_{\alpha} V^\alpha$ 
    & $\sim -T^A(z) - \overline{T}^A(\overline z) -(A\leftrightarrow B) + \text{const.}$ & $2$\\
$E$ & $\sum_{\alpha} (-1)^\alpha V^\alpha$ 
    & $\sim \epsilon^A(z,\overline z) + \epsilon^B(z,\overline z)$ & $4/5$\\
$U$ & $\sum_{\alpha} i^\alpha V^\alpha$ 
    &  $\sim$ linear comb. of $\Phi^{A,B}_{X\overline{\epsilon}}$ and $\Phi^{A,B}_{{\epsilon} \overline X}$ & $9/5$ \\
$\overline U$ & $\sum_{\alpha} (-i)^\alpha V^\alpha $ 
    & $\sim$ linear comb. of $\Phi^{A,B}_{X\overline{\epsilon}}$ and $\Phi^{A,B}_{{\epsilon} \overline X}$ &$9/5$  \\
\hline
\end{tabular}
\end{center}

\subsection{First and Second Order Perturbation Theory}

We consider the Hamiltonian
\[
H = H_0 + \lambda (H_1 + H_2), \qquad 0<\lambda \ll 1.
\]

At each half-integer site $j+\tfrac{1}{2}$, the decoupled term of the unperturbed Hamiltonian $H_0$ has a ground state 
\[
\ket{g} = \tfrac{1}{\sqrt{2}} (0,1,-1),
\] 
and two excited states $\ket{e_\pm}$ with energies $E_{e_\pm} = 1\pm \sqrt{3}$. They have Hadamard charges $i,-1$ and $1$ respectively.
The projector onto the low-energy subspace is
\[
P = \prod_j P_{j+\tfrac{1}{2}}, \qquad P_{j+\tfrac{1}{2}} = \ket{g}\bra{g}_{\,j+\tfrac{1}{2}}.
\]

The first order correction immediately evaluates to Potts$^2$.

\[
\lambda P V P =  \lambda \sum_{j,\alpha}   \left(  P\sigma^\alpha_{j+\frac{1}{2}} P\right)V^\alpha_{j+\frac{1}{2}} 
 = -\frac{\lambda}{2} \sum_j T_{j+\frac{1}{2}} \]

For higher order corrections, as per the Schrieffer-Wolff procedure ~\cite{Bravyi_2011,SW_1966} we use the  resolvent:

\[
\Gamma  = \sum_j G_{j+\frac{1}{2}}  + \text{multi-site excitations} \qquad G_{j+\frac{1}{2}} = \left( \sum_{\pm} \frac{\ket{e_{\pm}}\bra{e_{\pm}}}{E_g -E_{e_\pm}}   \right)_{j+\frac{1}{2}} \prod_{k \neq j} P_{k+\frac{1}{2}}
\]

Up to third order in perturbation theory we will only need the local resolvents $G_{j+\frac{1}{2}}$.
For each configuration of auxiliary spins $\lbrace s_{j+\frac{1}{2}}\rbrace$, we can define a Hilbert space $\mathcal{H}^{\lbrace s_{j+\frac{1}{2}}\rbrace}$. 
Then $V$ only has matrix elements between  $\mathcal{H}^{\lbrace s_{j+\frac{1}{2}}\rbrace}$ and  $\mathcal{H}^{\lbrace s'_{j+\frac{1}{2}}\rbrace}$ when the auxiliary-site configurations $\lbrace s_{j+\frac{1}{2}}\rbrace$ and $\lbrace s'_{j+\frac{1}{2}}\rbrace$ differ from each other on at most one site.

The $\mathcal{O}(\lambda^2)$ Schrieffer Wolff correction considers the contributions from single excitations on sites $j+\frac{1}{2}$. The terms generated are

\begin{equation}
\begin{split}
\lambda^2 P V \Gamma V P
=
\lambda^2 \sum_j \bra{g}V_{j+\frac{1}{2}} G_{j+\frac{1}{2}} V_{j+\frac{1}{2}} \ket{g}
\end{split}
\end{equation}

Using the aforementioned notation we can write each local contribution as

\[ \sum_{\alpha, \beta} V^{\alpha}_{j+\frac{1}{2}} V^{\beta}_{j+\frac{1}{2}} \bra{g} \sigma^\alpha G^{}_{j+\frac{1}{2}} \sigma^\beta \ket{g}  \]

The Hadamard symmetry leads to the following simplification:

\[ \sum_{\alpha, \beta} V^{\alpha} V^{\beta} \bra{g} \sigma^\alpha G^{}_{2j+1} \sigma^\beta \ket{g}  =   \sum_{\alpha, \beta, \pm }\left( (\pm i)^{\alpha-\beta} \bra{g} \sigma^{(0)} \frac{\ket{e_{\pm}}\bra{e_{\pm}}}{E_g - E_{e_\pm}} \sigma^{(0)} \ket{g} \right) V^{\alpha} V^{\beta} \]

From this we can directly see that each of the two `factors'' can only contribute a charge-conjugation odd operator (i.e. $V^{\alpha} - V^{\alpha+2}$), and any $\mathcal{O}(\lambda^2)$ perturbation term is a product of two such charge-conjugation odd operators , either on the same Potts chain or on different Potts chains. This directly rules out any term with the same symmetries as $\epsilon^A \epsilon^B$. In fact expanding directly gives

\begin{equation} \frac{1}{4}\lbrace U, \overline U\rbrace + \frac{\sqrt{3}}{8}[U,\overline{U}] \end{equation} 

Decomposing $U=U_A+U_B, \overline{U} = \overline{U}_A + \overline{U}_B$ on the two chains, we can work out the following contributions:

\begin{enumerate}
    \item $\frac{1}{4}\lbrace U_A ,\overline U_A\rbrace = \frac{1}{2}T_A-2$
    \item $ \lbrace U_A , \overline U_B\rbrace+ (A\leftrightarrow B)  \sim  \Phi^A_{X\bar \epsilon} \Phi^B_{\epsilon \bar X} + \Phi^B_{X\bar \epsilon} \Phi^A_{\epsilon \bar X}$
    \item $ \frac{\sqrt{3}}{8}[U_A,\overline U_A] = \frac{3}{4}\left((Y+W)_jZ^\dagger_{j+1} + \textrm{h.c.} \right)$
\end{enumerate}

The first contribution rescales the unperturbed Hamiltonian. The second contribution is highly irrelevant with dimension $18/5$. The third contribution can be identified via symmetries (Krammers-Wannier, charge conjugation, parity and time reversal) with $\partial_x \left(\epsilon^A(z,\overline z) - \epsilon^B(z, \overline z)\right)$; although it has dimension $9/5$ as confirmed by numerical correlation scaling, it is a total derivative in the field theory and does not gap out the CFT.

\subsection{Third Order Perturbation Theory}

We now check whether the $\epsilon^A \epsilon^B$ perturbation appears at third order.  
The correction takes the form
\[
    PV\Gamma V\Gamma VP - \tfrac{1}{2}\lbrace PVP,PV\Gamma^2 VP\rbrace .
\]

\noindent
We can isolate the auxiliary and dynamical degrees of freedom as follows (for notational simplicity we replace $j+\frac{1}{2}$ with $j$ in subscripts):
\[
\begin{aligned}
   &\left((P\sigma^\alpha G \sigma^\beta G \sigma^\gamma P)
   - \tfrac{1}{2} (P\sigma^\alpha P)(P\sigma^\beta G^2 \sigma^\gamma P)
   - \tfrac{1}{2} (P\sigma^\alpha G^2 \sigma^\beta P)(P\sigma^\gamma P) 
   \right)V^\alpha_j V^\beta_j V^\gamma_j \\
   &\quad + \tfrac{1}{2} \sum_{j \neq k} (P\sigma^\alpha G^2 \sigma^\gamma P)(P\sigma^\beta P)
   \left( [V_j^\alpha, V_k^\beta] V_j^\gamma + V_j^\alpha [V_j^\beta, V_k^\gamma] \right).
\end{aligned}
\]

Notice that any factor $(\sum_{\alpha} P\sigma^\alpha P )V^\alpha$ is porportional to $T$, while factors of the form $(P\sigma^\alpha G^2 \sigma^\beta P) V^\alpha V^\beta$ are products of two charge-conjugation-odd operators. The $\epsilon^A \epsilon^B$ perturbation cannot arise from any of the terms containing these factors.
Thus it suffices to focus on the first term:

\begin{equation}
\begin{split}
  C_{\alpha, \beta, \gamma} V^\alpha_j V^\beta_j V^\gamma_j \qquad  C_{\alpha,\beta,\gamma} \equiv  \bra{g} \sigma^\alpha G \sigma^\beta G \sigma^\gamma \ket{g}
\end{split}
\end{equation}

The coefficient evaluates to

\[
C_{\alpha,\beta,\gamma} = C_{\alpha+1,\beta+1,\gamma+1}, \qquad
C_{\alpha,\beta,\gamma} =
\begin{cases}
\dfrac{(-1)^{m+n}}{8}, & (\alpha,\beta,\gamma)=(2m,0,2n), \\[6pt]
0, & (\alpha,\beta,\gamma)=(2m+1,0,2n+1), \\[6pt]
\pm\dfrac{i}{8\sqrt{3}}, & (\alpha,\beta,\gamma)=(m,0,m\pm1).
\end{cases}
\]

Then $\sum_{\alpha, \beta,\gamma} C_{\alpha,\beta,\gamma}  V^\alpha V^\beta V^\gamma$ takes the form

\begin{equation}
    \frac{1}{32} \left( U T \overline U + \overline U T U  + U E U + \overline U  E \overline U\right)  +\frac{1}{16\sqrt{3}}\left(  \overline U  T U - U T \overline U \right)
\end{equation}

Each term is has three factors. For a perturbation coupling the two chains, two of the factors must belong to the $A$ chain and one of them to the $B$ chain or vice-versa. Recalling that $U$ and $\overline U$ are charge-conjugation-odd, a perturbation such as $\epsilon^A \epsilon^B$ can only arise when the middle factor is on $A$ chain and the $U, \overline{U}$ factors are both on the $B$ chain or vice versa. The only such contribution is
$U_AE_BU_A + \overline U_A E_B \overline U_A + (A\leftrightarrow B)  = 4E_A E_B$. 
We conclude that the relevant perturbation arises at third order with coefficient $\lambda^3/8$.

\end{document}